\documentclass[twocolumn, longauth, ]{aa}

\usepackage{graphicx}
\usepackage{txfonts}
\usepackage{xcolor}
\usepackage{natbib}
\usepackage{subfig}
\usepackage{afterpage}
\usepackage{color}
\usepackage{siunitx}
\usepackage{xcolor}
\usepackage{afterpage}
\usepackage{amsmath}
\usepackage[colorlinks=true,linkcolor=blue,citecolor=blue,urlcolor=blue]{hyperref}

\newcommand{\Reff}{$R_{\rm eff}$}

\newcommand{\Vsyst}{$V_{\rm syst}$}

\newcommand{\Vlos}{$V_{\rm LOS}$}
\newcommand{\Slos}{$\sigma_{\rm LOS}$}
\newcommand{\PaperI}{\hyperlink{cite.Iodice2023}{Paper~I}}
\newcommand{\PaperII}{\hyperlink{cite.Buttitta2025}{Paper~II}}
\newcommand{\PaperV}{\hyperlink{cite.Doll2026}{Paper~V}}

\begin{document}

   \title{Looking into the faintEst WIth MUSE (LEWIS): Exploring the nature of ultra-diffuse galaxies in the Hydra-I cluster}

   \subtitle{VI. A star-forming UDG in Hydra~I: a rare UDG or a transition phase?}

   \author{Luca Rossi\inst{1,2}\fnmsep
          \and Chiara Buttitta\inst{1}\thanks{\email{chiara.buttitta@inaf.it} }
          \and Goran Doll\inst{1,2}
          \and Enrichetta Iodice\inst{1}
          \and Marco Gullieuszik\inst{3}
          \and Marc Sarzi\inst{4}
          \and Marco Mirabile\inst{5,6}
          \and Johanna Hartke\inst{7,8,9}
          \and Magda Arnaboldi\inst{10}
          \and Rosa Calvi\inst{1}
          \and Michele Cantiello\inst{11}
          \and Enrico Maria Corsini\inst{3, 12}
          \and Giuseppe D'Ago \inst{13}
          \and Jes{\'u}s Falc{\'o}n-Barroso \inst{14,15}
          \and Francesca Fonzo\inst{1}
          \and Duncan A. Forbes \inst{16}          
          \and Michael Hilker\inst{10}
          \and Antonio La Marca\inst{17,18}
          \and Alessandro Loni\inst{1}
          \and Steffen Mieske \inst{19}  
          \and Maurizio Paolillo \inst{1,2}
          \and Marina Rejkuba \inst{10}
          \and Marilena Spavone \inst{1}
          \and Chiara Spiniello \inst{10, 20, 1}
          }
          
        \institute{INAF $-$ Astronomical Observatory of Capodimonte, Salita Moiariello 16, I-80131, Naples, Italy
        \and
        University of Naples ``Federico II'', C.U. Monte Sant'Angelo, Via Cinthia, 80126, Naples, Italy
        \and
        INAF $-$ Osservatorio Astronomico di Padova, Vicolo dell’Osservatorio 5, I-35122 Padova, Italy
        \and
        Armagh Observatory and Planetarium, College Hill, Armagh, BT61 9DG UK
        \and
        INAF $-$ Astronomical Observatory of Abruzzo, Via Maggini, 64100, Teramo, Italy
        \and
        Gran Sasso Science Institute, viale Francesco Crispi 7, I-67100 L'Aquila, Italy
        \and
        Finnish Centre for Astronomy with ESO, (FINCA), University of Turku, FI-20014 Turku, Finland
        \and
        Tuorla Observatory, Department of Physics and Astronomy, University of Turku, FI-20014 Turku, Finland
        \and
        Turku Collegium for Science, Medicine and Technology (TCSMT), University of Turku, FI-20014 Turku, Finland
        \and
        European Southern Observatory, Karl$-$Schwarzschild-Strasse 2, 85748 Garching bei München, Germany
        \and
        INAF $-$ Astronomical Observatory of Abruzzo, Via Maggini, 64100, Teramo, Italy
        \and
        Dipartimento di Fisica e Astronomia ``G. Galilei'', Universit\`a di Padova, vicolo dell'Osservatorio 3, I-35122 Padova, Italy
        \and
        Institute of Astronomy, University of Cambridge, Madingley Road, Cambridge CB3 0HA, UK 
        \and
        Instituto de Astrofísica de Canarias, Calle V\'ia L\'actea s/n, 38200 La Laguna, Spain
        \and
        Departamento de Astrof\'isica, Universidad de La Laguna, E-38200, La Laguna, Tenerife, Spain
        \and
        Centre for Astrophysics \& Supercomputing, Swinburne University of Technology, Hawthorn VIC 3122, Australia
        \and
        European Space Agency/ESTEC, Keplerlaan 1, 2201 AZ Noordwijk, The Netherlands
        \and
        Leiden Observatory, Leiden University, PO Box 9513, NL-2300 RA Leiden, The Netherlands
        \and
        European Southern Observatory, Alonso de Cordova 3107, Vitacura, Santiago, Chile
        \and
        Sub-Dep. of Astrophysics, Dep. of Physics, University of Oxford, Denys Wilkinson Building, Keble Road, Oxford OX1 3RH, United Kingdom}

\date{Received MM DD, YYYY; accepted MM DD, YYYY}

  \abstract  
   {This paper presents a detailed analysis of a gas-rich star-forming ultra-diffuse galaxy (UDG) as part of the ESO Large Programme 'Looking into the faintEst WIth MUSE (LEWIS)'. Among the UDGs in the LEWIS sample, UDG\,6 is the only galaxy that hosts a significant amount of ionised gas with evidence of emission lines, suggesting recent star-forming activity.}
   {The main goal of this work is to constrain the formation history of this UDG by comparing its properties with the main formation scenarios proposed for this extreme class of galaxies.}
   {We adopted integral field spectroscopy from MUSE to derive the morphology and the structural properties of the stellar and gas components of UDG\,6. We applied spectral fitting and Voronoi tessellation algorithms to the MUSE data-cube to derive the kinematics and properties of the gas and stellar component, explore gas dynamics, and characterise emission lines. Moreover, we derived the GCs populations' properties by applying a multi-band spectrophotometric analysis.}
   {We confirmed that UDG\,6 is a member of Hydra I cluster with a systemic velocity of $V_{\rm sys, \ast}\sim3580$\,km s$^{-1}$. It is characterised by a regular and elongated shape and contains a significant dust content ($A_V=1.2\,\pm\,0.5$ mag), a metal-poor ($12+\log_{10}(\text{O/H})=7.7\,\pm\,0.2$ dex) ionised gas fraction ($f_{\rm gas}=0.24\,\pm\,0.08$) and an underlying old-to-intermediate ($\gtrsim3$\,Gyr) stellar component. Evidence of local and clumpy star-forming activity has been revealed through the analysis of emission line ratios, and an arc-like tidal feature was discovered from unsharp masking analysis. The number counts of GCs in UDG\,6 is $N_{\rm GC}= 0.2 \pm 5.4$.} 
   {Like the other UDGs in LEWIS, UDG\,6 might originate from a `puffed-up dwarf' whose stellar content has been stretched out to larger radii, passively evolving into a more diffuse galaxy. Being located in a dynamically active region of the cluster, characterised by tidal features and stripping phenomena, we suggest that the environmental processes have played a role in shaping the properties of UDG\,6. A tidal interaction with a nearby galaxy might have triggered recent star-formation activity, without dramatically altering the coherent gas rotation in UDG\,6.}

\titlerunning{VI. The case study of UDG\,6}
\authorrunning{L. Rossi et al.}
\maketitle

\section{Introduction}
\label{sec:intro}
Ultra-diffuse galaxies are low-mass ($M_\ast\sim10^{5-9}\,$ M$_\odot$) galaxies 
empirically defined to have a central surface brightness of $\mu_{0,g}>24$\,mag arcsec$^{-2}$ 
and effective radius of \Reff\,>\,1.5\,kpc \citep{vanDokkum2015}. 
They are generally considered to represent the extreme low-surface-brightness 
(LSB) extension of the dwarf galaxy population.

Large populations of UDGs have been mainly identified and studied in dense environments such as clusters
of galaxies and groups, where the majority of deep and wide optical imaging surveys focused 
\citep{VanDerBurg2016, Roman2017, Spekkens2018, ManceraPina2019, Iodice2020, Lim2020, Kado-Fong2022, LaMarca2022b, 
Zaritsky2023udgs, Marleau2025, Makda2025}. UDGs in moderate-to-low density environments, such as filaments and isolated 
fields, have also been detected via HI surveys \citep{Trujillo2017, Leisman2017, 
ManceraPina2020, Janowiecki2019, Karunakaran2020, Adams2026}.
The dataset of UDGs collected so far allows us to identify several classes of UDGs with 
distinct properties suggesting different formation channels \citep[see ][for a review]{Gannon2026}.

UDGs in dense environments are typically red, quenched, and gas-poor. Their metallicity,
globular clusters (GCs), and dark matter (DM) content span wide ranges of values, which correlate
with the environment where they reside and the physical mechanisms they underwent in the past
\citep{Gannon2022, FerreMateu2023, Buzzo2024, Buttitta2025, Doll2026}. UDGs located in groups, in 
clusters' outskirts or in fields, are instead bluer and star-forming, dusty and gas-rich, with an 
irregular shape and extended and coherent velocity fields \citep{Prole2019, ManceraPina2020, Gault2021}. 

Despite the existence of multiple classes of UDGs being largely accepted by the scientific community, the present-day picture of the UDG framework remains mostly unclear because the current dataset of UDGs
in different environments is unbalanced. While deep optical imaging surveys have identified 
thousands of UDG candidates by detecting their stellar component (e.g \citealp{Zaritsky2023}), 
their neutral or ionised gas counterpart has been revealed only for a small number of UDGs 
(see e.g, \citealp{Kadowaki2017, Leisman2017, Papastergis2017, Janowiecki2019}). 
Therefore, it remains unclear whether the observed environmental differences between UDG populations 
reflect a genuine physical process or are partly driven by observational selection effects. 
Cluster UDGs could be the last evolutionary stage of the same type of objects located outside the clusters
\citep{Roman2017}. Once they enter dense environments, mechanisms such as ram-pressure stripping, 
tidal interactions, and galaxy harassment can efficiently remove gas from galaxies, potentially 
transforming gas-rich, star-forming UDGs into quiescent ones \citep{Venhola2017, Wittmann2017,Safarzadeh2017, ManceraPina2018, Sales2020}.

Spectroscopic observations are fundamental to fill the gap between stellar and gas counterpart statistics. 
By probing the kinematics, gas content, and emission lines, spectroscopy provides direct insight 
into ongoing star formation, stellar populations, and the impact of environmental mechanisms. These 
quantities are key parameters for testing how different formation channels produce inherently 
distinct UDGs populations. In this context, the 'Looking into the fainEst WIth MUSE' 
(LEWIS) project provided a precious contribution to improving our understanding of UDGs 
\citep{Iodice2023}. Combining deep optical data from VST and integral-field (IF) 
from the MUSE spectrograph, LEWIS revealed the existence of multiple classes of UDG population
in a single environment, characterised by different properties \citep{Buttitta2025, Doll2026}.

In this paper, we present a detailed analysis of UDG\,6, a rare example of a gas-rich UDG in Hydra, as evidenced by clear emission lines from ionised gas. The paper is organised as follows.  In Section~\ref{sec:observations} 
we describe the structural properties and morphology of UDG\,6 and describe the spectroscopic data 
used throughout this work. In Section~\ref{sec:photometry}, we show the morphological analysis 
performed on the MUSE reconstructed images. In Section~\ref{sec:spectroscopy}, we 
discuss the spectroscopic analysis performed on the MUSE data-cube to derive properties 
of stellar and gas components in UDG\,6. In Section~\ref{sec:discussion}, we discuss the physical 
implications of the obtained results within the environment and in the LEWIS sample framework.
Finally, in Section~\ref{sec:conclusions}, we report our conclusions and future perspectives.
Throughout the paper, we adopt the standard $\Lambda$CDM cosmology \citep{Hinshaw2013}: 
$H_0=70$\,km\,s$^{-1}$\,Mpc$^{-1}$, $\Omega_{\rm m}=0.3$, and $\Omega_\Lambda=0.7$.

\section{Observations}
\label{sec:observations}

\begin{figure*}[!h]
    \centering
    \includegraphics[scale=0.61, trim={1.3cm 0cm 0cm 0cm}, clip]{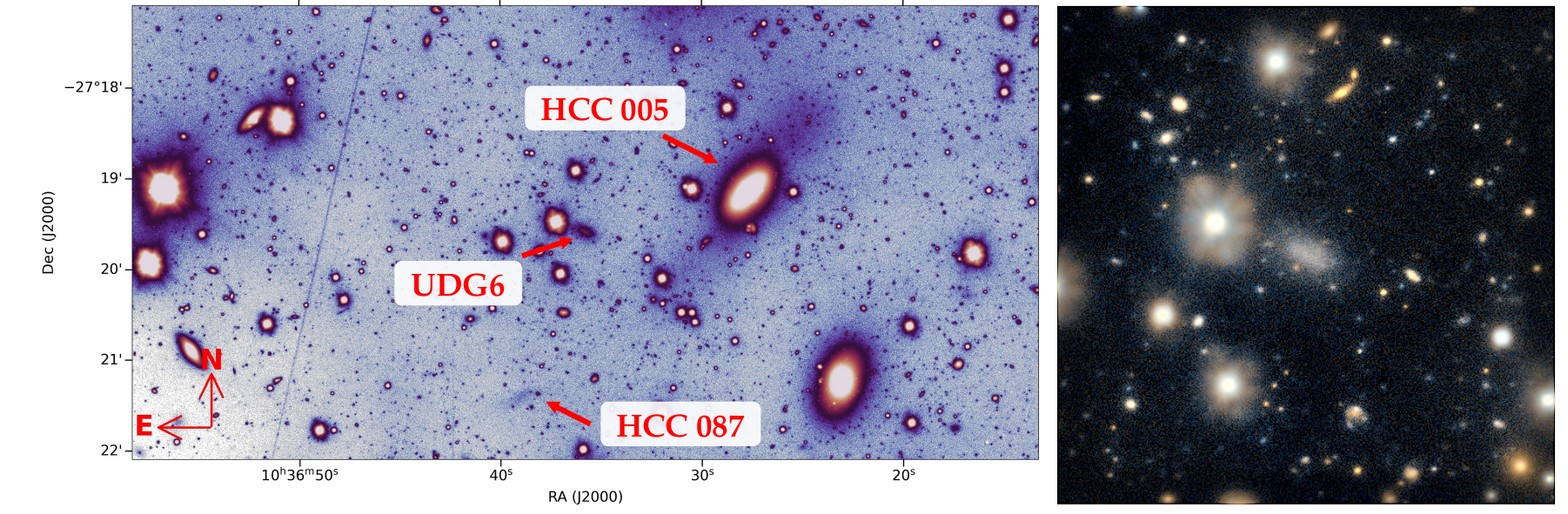}
    \caption{OmegaCAM@VST $r$-band image of the North group of the Hydra I cluster, adapted from ESO/INAF/M. Spavone, E. Iodice. Left panel: The field-of-view is 10\,$\times$\,5\,arcmin$^2$. Right panel: Zoom-in $g-r$ colour-composite image of UDG\,6. The field-of-view is 1.6\,$\times$\,1.6\,arcmin$^2$}
    \label{fig:UDG6_VST}
\end{figure*}

\subsection{Properties of UDG\,6 from deep imaging}
UDG\,6 was originally identified in the VST Early-type Galaxy Survey 
(VEGAS\footnote{for details, see 
\href{https://sites.google.com/inaf.it/vegas/home}{https://sites.google.com/inaf.it/vegas/home}},
\citealt{Capaccioli2015, Iodice2021b}) performed 
at the European Southern Observatory (ESO) VLT Survey Telescope (VST). 
VEGAS is a deep multi-band imaging survey of nearby (z\,<\,0.05) groups and 
clusters of galaxies, to map the light distribution down to a surface 
brightness level of $\mu_g\sim27-28$\,mag arcsec$^{-2}$ in $g$ band. 

UDG\,6 is located in the Hydra I cluster of galaxies, a dense 
environment of galaxies at a distance $D\sim51$\,Mpc \citep{Christlein2003} 
which shows signs of ongoing interactions and mass assembly around 
the dominant central cluster galaxy NGC3311 \citep{Arnaboldi2012, Spavone2024, Hess2022}, 
and it belongs to the first sample of UDGs discovered in Hydra 
I \citep{Iodice2020}. UDG\,6 is located at a projected cluster-centric
distance of 12.2 arcmin from the cluster centre\footnote{We assumed 
as cluster centre the coordinates of the bright cluster member NGC3311 
(RA = 159.17842$^\circ$, Dec = -27.528339$^\circ$) as done in previous 
works \citep{Forbes2023,Buttitta2025}.},  corresponding to a 
projected physical distance of {$\sim180\,{\rm kpc}$ and belongs to the north 
group of galaxies in Hydra~I (Fig.\ref{fig:UDG6_VST}, 
left panel). UDG\,6 is in proximity to the lenticular galaxy HCC\,005, which shows 
two prominent tidal tails, signs of the presence of tidal interactions 
acting on the galaxy \citep{LaMarca2022a, Spavone2024}. 
The morphology of UDG\,6 is regular, characterised by an elongated 
spheroidal shape (Fig.\ref{fig:UDG6_VST}, right panel). Among the UDGs in Hydra I 
\citep{Iodice2020, LaMarca2022a}, UDG\,6 is one of the bluest 
($g-r = 0.32$\,mag) and least massive ($M_\ast=3.2\cdot10^7$\,M$_\odot$) 
of the sample. The stellar mass estimate has been 
obtained from photometric analysis, by using the galaxy's 
absolute magnitude, colour, and the relation presented in \cite{Into2013}. 
Being very blue, the stellar mass could be a lower limit.
We used deep $H$-band data from VIRCAM@VISTA telescope 
(Prog. ID. 109.231E.00, P.I. M. Cantiello) to derive an independent estimate
of the stellar mass, and we obtained a value of $M_\ast=(1.8 \pm 0.72) \cdot10^7$\,M$_\odot$,
consistent with the previous estimate within 2$\sigma$. 
Finally, UDG\,6 does not seem to have any GC candidates \citep{Iodice2020}.
In Table~\ref{tab:properties}, we report the structural parameters for UDG\,6.

\begin{table}
\centering
\caption{Structural properties of UDG\,6.} 
\renewcommand{\tabcolsep}{0.35cm}
\renewcommand{\arraystretch}{1.0}
\begin{tabular}{l c l  c }
\hline
\hline
     & Property &                     & UDG\,6               \\
\hline
(1)  & RA       & [J2000]             & +10h\,36m\,35.80s  \\
(2)  & Dec      & [J2000]             & -27d\,19m\,36.12s  \\
(3)  & $M_r$    & [mag]               & -14.38 $\pm$ 0.08  \\
(4)  & $g-r$    & [mag]               & 0.32 $\pm$ 0.20    \\
(5)  & $\mu_0$  & [mag arcsec$^{-2}$] & 24.08 $\pm$ 0.13   \\
(6)  & \Reff    & [kpc]               & 1.37 $\pm$ 0.12    \\
(7)  & $M/L$    & [M$_\odot$/L$_\odot$] & 0.70               \\  
(8)  & $M_\ast$ & [$10^8$\,M$_\odot$]   & 0.32               \\
\hline 
\end{tabular}
\tablefoot{Galaxy properties of UDG\,6 from \cite{Iodice2020}. (1-2): the Right Ascension and Declination of the galaxy. (3): absolute magnitude in the $r$ band. (4): average $g-r$ colour. (5-6): central surface brightness and effective radius, respectively. (7-8): stellar mass-to-light ratio and stellar mass derived from $r$ band, respectively.}
\label{tab:properties}
\end{table}

\subsection{Spectroscopic data from LEWIS}

Spectroscopic data of UDG\,6 were acquired under the LEWIS\footnote{for details, see \href{https://sites.google.com/inaf.it/lewis/home}{https://sites.google.com/inaf.it/lewis/home}} project (Prog. ID. 108.222P, P.I. E. Iodice), whose observations 
were carried out at the ESO Multi Unit Spectroscopic Explorer (MUSE, \citealt{Bacon2010}). 
Configured in wide-field mode, MUSE offers a field-of-view (FOV) of 
1\,$\times$\,1\,arcmin$^2$ with a spatial sampling of 0.2\,arcsec~pixel$^{-1}$.
The MUSE nominal wavelength range is $4800-9300$\,\AA, with a spectral sampling 
of $1.25$\,\AA~pixel$^{-1}$ and an average spectral resolution of 
FWHM$~=2.51$\,\AA\,\citep{Bacon2017}. The spectral resolution of 
the LEWIS data was derived in \citealt{Buttitta2025} (hereafter \PaperII). It varies from FWHM=2.75\,\AA\,
($\sigma\sim70$\,km s$^{-1}$) at $\lambda\sim5000$\,\AA\,to FWHM=2.63\,\AA\,($\sigma\sim37$\,km s$^{-1}$) at $\lambda\sim9000$\,\AA.

Data of UDG\,6 were acquired on three different observation blocks (OB): $23-24$ April 2023, 
$18-19$ December 2023, and $12-13$ January 2024. Within each OB, three exposures, with an integration 
time of 900 seconds for each, were obtained. The total exposure time on target is 2.25 hours,
with an average seeing of FWHM $\sim1.09$\,arcsec. A standard data reduction 
was performed using the MUSE pipeline \citep{Weilbacher2020} 
running on the {\sc esoreflex} workflow \citep{Freudling2013}. As for other 
galaxies in LEWIS, we used the previous result to perform a second improved data reduction,
following the procedures described in \PaperII, to improve the data quality and 
reduce sky-background fluctuations. MUSE data have been used to perform a 
detailed analysis of the galaxy's morphology and properties.

\section{Morphology of UDG\,6}
\label{sec:photometry}

\subsection{Isophotal analysis}
\label{sec:isophotal_analysis}

\begin{figure*}[!t]
    \centering
    \includegraphics[scale=0.35]{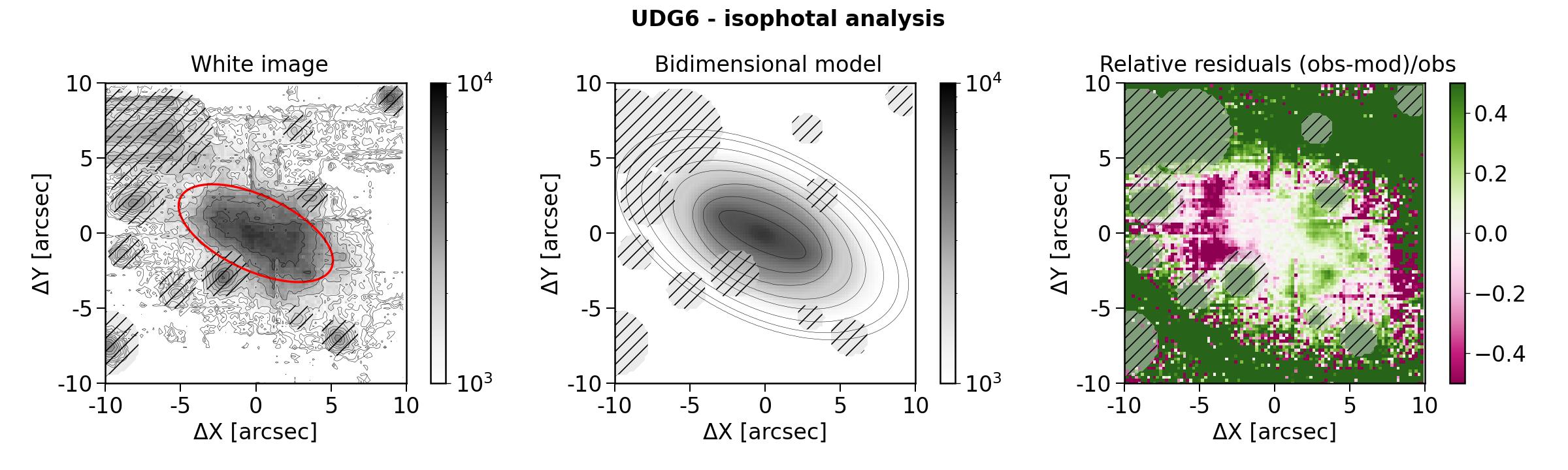}
    \includegraphics[scale=0.35]{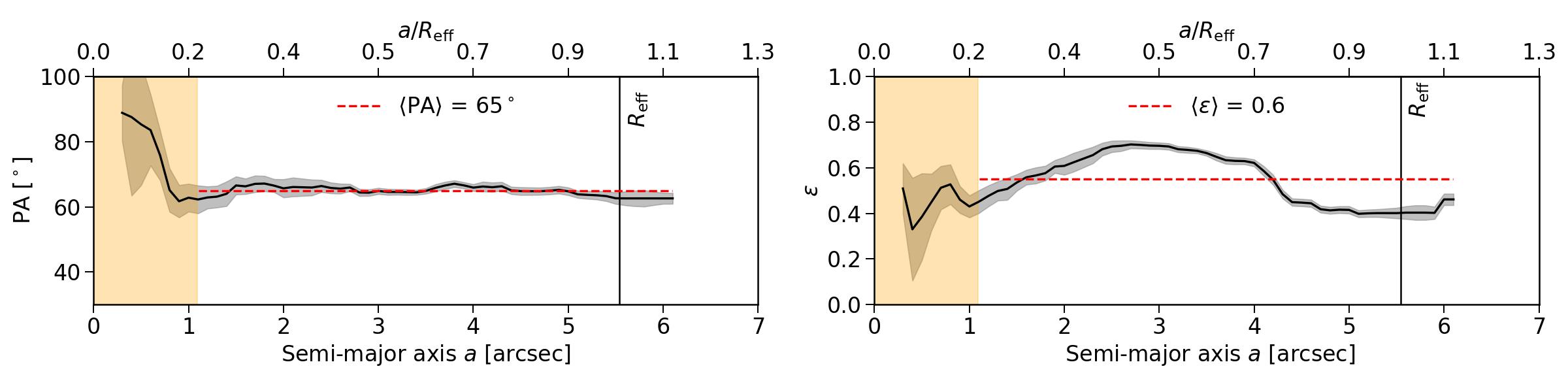}
    \caption{Isophotal analysis of UDG\,6. Top panels: MUSE reconstructed white image of UDG\,6 with contours (left), bi-dimensional model from isophotal fitting (centre), and relative residuals (right). The grey hatched circles represent the masked regions. The red ellipse represents the position of \Reff\ and marks the region used to extract the stacked spectrum. Bottom panels: Radial profiles of PA (left) and $\epsilon$ (right) of fitted isophotes (solid lines) with $1\sigma$ uncertainties (grey shaded regions). The orange shaded region marks the PSF-limited region. The mean geometric parameters of the galaxy are marked by red dashed lines. The vertical black line marks the position of \Reff.}
\label{fig:isophotal_analysis}
\end{figure*}

We recovered the mean geometric parameters of UDG\,6 by performing an 
isophotal analysis on the reconstructed MUSE white image 
(Fig.~\ref{fig:isophotal_analysis}, left panel) using the {\sc ellipse} 
task of the Python package {\sc photutils} \citep{photutils}. 
This type of analysis has already been carried out on deep VST imaging 
data \citep{Iodice2020}. However, we will take advantage of the wider 
spectral coverage of MUSE to capture the light contribution of both 
stellar and gas components in UDG\,6.

We followed the prescription described in \PaperII: we masked 
all the foreground stars, background galaxies, and spurious sources
in the FOV, then we performed a first isophotal fit to determine the centre
of the galaxy. Therefore, we allowed the centre coordinates of the ellipses free to vary 
and fixed the values of position angle (PA) and ellipticity ($\epsilon$)
to the values obtained from the isophotal analysis performed on VST 
images. We iteratively computed the mean and standard deviation of 
centre coordinates values of fitted isophotes to identify the radial 
convergence interval, i.e. the radial range where $\langle x\rangle$ 
and $\langle y \rangle$ reach stability within $\sigma_x$ and $\sigma_y$, 
respectively. The values of $\langle x\rangle$ and $\langle y \rangle$ 
computed in the radial region ${\rm FWHM} < a \lesssim 0.3\,R_{\rm eff}$, 
with $a$ semi-major axis of the fitted isophote and $R_{\rm eff}$ effective 
radius \citep{Iodice2020}, determined the galaxy centre $(x_c, y_c)$. 
This preliminary analysis revealed a shift in the centre coordinate, 
with respect to the previously estimate, towards 
the eastern direction of $\Delta x\sim1\,$arcsec beyond 
$\gtrsim 0.4\,R_{\rm eff}$, probably due to a possible light
contamination from the nearby bright star.

We performed a second fit of the galaxy isophotes by fixing 
the centre coordinates previously defined 
and letting the $\epsilon$ and PA of the isophotes free to vary.
By inspecting the radial profiles of the fitted isophote parameters (Fig.~\ref{fig:isophotal_analysis}, bottom panels),
we found that the orientation of the fitted isophotes remains nearly 
constant, while the ellipses become more 
elongated with increasing radial distance, reaching a peak of $\epsilon\sim0.7$ at 
$a\sim0.5\,R_{\rm eff}$ and becoming more roundish again at $a\sim1\,R_{\rm eff}$. 
We obtained the geometric parameters of the galaxy
by computing the average values of the PA and $\epsilon$ of the best-fitted 
isophotes in the radial range between the FWHM of the PSF ($\sim0.2\,R_{\rm eff}$)
and the largest radius where the isophotes best-fit is reliable ($\sim1.1\,R_{\rm eff}$).} 
We therefore obtained $\langle\epsilon\rangle=0.6$ and $\langle$PA$\rangle$\,$=65^\circ$,
consistent with the values obtained from isophotal analysis 
performed on VST images \citep{Iodice2020}. We used these 
parameters ($x_c, y_c, \langle \epsilon \rangle, \langle {\rm PA} \rangle$, $R_{\rm eff}$) 
to define the elliptical aperture within which we co-added all 
the spaxels and derived the stacked spectrum, which was then 
used to perform the integrated spectral analysis of UDG\,6.

\subsection{Unsharp mask}
\label{sec:unsharp_mask}

\begin{figure*}[!t]
    \centering
    \includegraphics[scale=0.45]{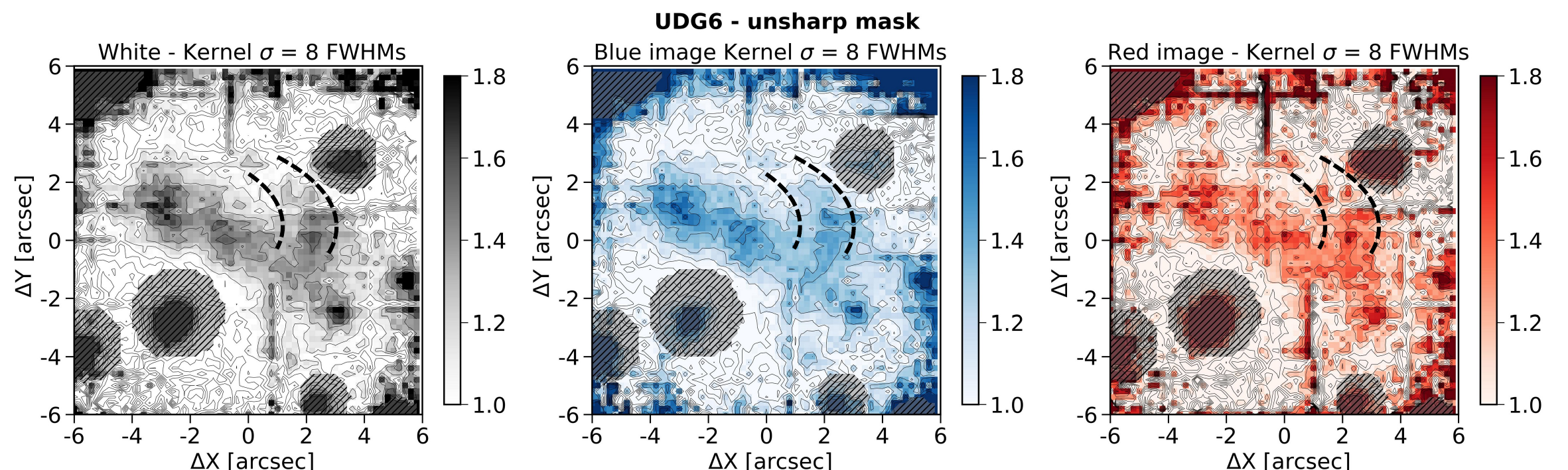}
    \caption{Unsharp mask of UDG\,6. Total white image ($4800<\lambda<9200$\,\AA, left), blue image ($4800<\lambda<7000$\,\AA, centre), and red image ($7000<\lambda<9200\,$\AA, right) obtained by collapsing the MUSE data-cube along the spectral direction. The adopted Gaussian kernel size is $\sigma_{\rm kernel}=8\,{\rm FWHMs}$. Grey hatched circles represent the masked regions, while the black dashed line delimits the spiral-like feature.}
\label{fig:unsharp_mask}
\end{figure*}

The isophotal analysis revealed a regular morphology in UDG\,6,
characterised by elongated isophotes with constant orientation 
on the sky plane. However, this type of analysis can give us 
hints only on the global symmetry and large-scale structures
of the galaxy. Therefore, we decided to perform an unsharp 
mask analysis on the MUSE reconstructed image to investigate 
the presence of small-scale features and sub-structures that 
could be hidden by the diffuse light distribution of UDG\,6. 
For this reason, we convolved the image with Gaussian kernels
of different sizes in units of the MUSE FWHM in the range
2-14\,FWHM to maximise the light contrast of possible 
features. Fig.~\ref{fig:unsharp_mask},
shows the unsharp mask on the white image of UDG\,6 obtained from 
the best kernel size ($\sigma_{\rm kernel} = 8\,{\rm FWHMs}$). 
We applied the same analysis on two additional images, obtained by collapsing the 
data-cube in the bluest ($4800<\lambda<7000$\,\AA, Fig.~\ref{fig:unsharp_mask}, central panel)
and in the reddest ($7000<\lambda<9200\,$\AA, Fig.~\ref{fig:unsharp_mask}, right panel) 
spectral range. The two ranges were chosen to split the spectral region 
where emission lines were visible, from the longer wavelength, more 
contaminated and without emissions. This multi-band approach aims to reveal not only the various 
coherent structures that constitute the galaxy, but also assess 
the relative light contribution of the different components 
dominant in certain spectral ranges.

In both the unsharp-masked white-light and blue-band images of UDG\,6, a distinct 
arc or spiral-like feature is visible, extending from the right side 
of the galaxy's main elongated body and curving towards the north-east 
(see Fig.~\ref{fig:unsharp_mask}).
The feature is less evident in the red image, which appears 
clumpier and disturbed. The red image also appears noisier since 
the redder spectral range of the MUSE data-cube contains several 
sky-line residual features. Since the blue image ($\lambda<7000$\,\AA) 
contains prominent gas emission lines (see Section~\ref{sec:kin_1D}), we can reasonably assume 
that the arc-like feature is associated with these strong emissions
rather than the underlying stellar continuum 
(see also Section~\ref{sec:spectroscopy}). These findings  
suggest a possible gravitational interaction 
or distortion on UDG\,6 due to the proximity of nearby lenticular
galaxy HCC\,005 (Fig.~\ref{fig:UDG6_VST}, left panel). We will present 
a more detailed discussion of this hypothesis in Section~\ref{sec:discussion}.

\section{Spectroscopic analysis}
\label{sec:spectroscopy}

In this section, we describe the spectroscopic analysis performed on UDG\,6. 
We recovered the integrated global properties (1D) such as kinematics 
(Section~\ref{sec:kin_1D}) and properties of stars (Section~\ref{sec:stellar_pop}), 
and gas (Sections~\ref{sec:star_forming_regions} and ~\ref{sec:gas_pop}), 
as well as, taking advantage of the IF nature of LEWIS data, spatially-resolved
(2D) gas kinematics maps (Section~\ref{sec:gas_vel_map}).

\subsection{Integrated stellar and gas kinematics}
\label{sec:kin_1D}

\begin{figure*}[!t]
    \centering
    \includegraphics[scale=0.28]{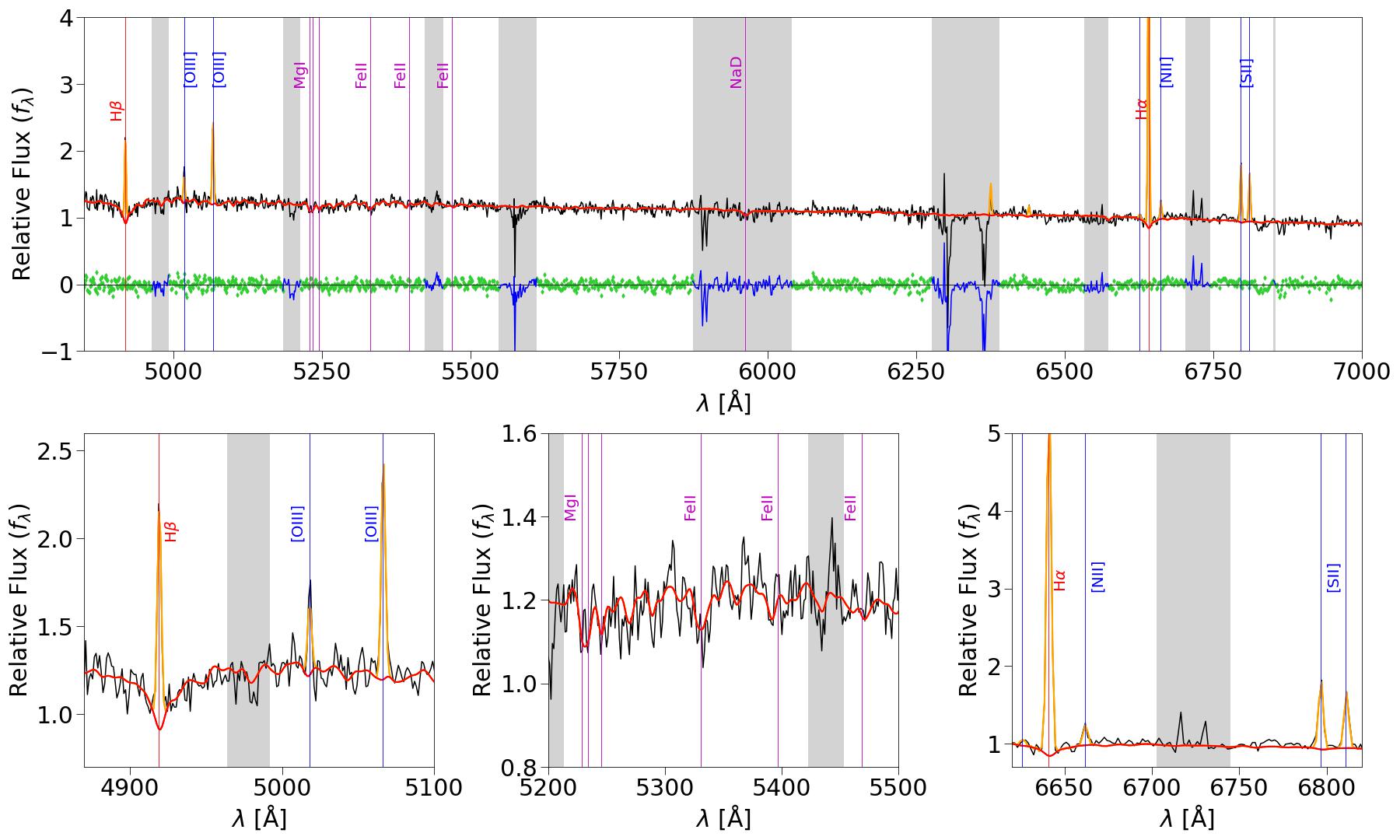}
    \caption{Stacked spectrum of UDG\,6 with its best-fit. Top row: the stacked spectrum is shown as a black line, and the best-fit obtained with {\sc pPXF} is shown as a red solid line for the stellar component and in yellow for the gas components. The main absorption lines are marked in magenta, the Balmer emission lines in red, and the forbidden ones in blue, respectively. The residuals between the observed spectrum and the best-fit are shown as green points. The grey areas are the masked spectral regions of the spectra and are excluded from the fit. Bottom row: enlarged regions around the main fitted absorption and emission lines of the top-row fit.}
    \label{fig:UDG6_bestfit}
\end{figure*}

The main peculiarity of UDG\,6 resides in the presence of strong 
emission lines (Fig.~\ref{fig:UDG6_bestfit}, top panel). 
Among the galaxies in LEWIS \citep{Iodice2023}, UDG\,6 is the only UDG which 
shows prominent emission lines, such as Balmer lines
\ion{H}{$\beta$} ($\lambda\,4861$\,\AA) and \ion{H}{$\alpha$} ($\lambda\,6563$\,\AA), 
the [\ion{O}{III}] doublet ($\lambda\lambda\,4959,5007$\,\AA), the reddest line of the [\ion{N}{II}] 
doublet ($\lambda\lambda\,6548, 6583$\,\AA)\footnote{the bluest one, which is always fainter, is just 
barely detectable.} and the [\ion{S}{II}] doublet ($\lambda\lambda\,6716,6731$\,\AA). 
While other UDGs in LEWIS show clear absorption features such 
as \ion{H}{$\beta$} and \ion{H}{$\alpha$}, 
magnesium (\ion{Mg\,I}{}, $\lambda\lambda\,5167,5173, 5184$\,\AA) and the
calcium (\ion{CaT}{}, $\lambda\lambda\,8498,8542,8662$\,\AA) triplets and iron 
absorption lines (\ion{Fe\,I}{}, $\lambda\lambda\,5198.71,5270.40, 5335.16$\,\AA), 
the spectrum of UDG\,6 only shows extremely weak stellar absorption features. 
The spectral profile of \ion{H}{$\beta$} shows hints of a broad absorption
line with emission on top (Fig.~\ref{fig:UDG6_bestfit}, bottom left panel).
A few extremely weak iron absorption lines are also present in the spectrum, while 
the \ion{Mg\,I}{} triplet is barely visible 
(Fig.~\ref{fig:UDG6_bestfit}, bottom central panel). 
At wavelengths longer than $\sim7000$\,\AA, the spectrum is strongly affected 
by residuals from sky emission lines, with the calcium triplet region 
(CaT; $\lambda \sim 8500$-$8800$\,\AA) being particularly noisy and heavily contaminated.
We carefully inspected this region, and the effect of the Zurich Atmosphere Purge algorithm
\cite[ZAP, ][]{Soto2016}, realizing that ZAP creates a fictitious absorption-line CaT 
(see \PaperII\, for a description of the adopted ZAP recipe). 
Therefore, we decided to focus our analysis on the optical spectral range up 
to 7000\,\AA, which has an average FWHM$\,\sim\,2.69\,$\AA\, corresponding to $\sigma\approx49\,$km s$^{-1}$.

Following the same approach presented in \PaperI\,and \PaperII, 
we performed spectral fitting on the stacked spectrum obtained from 
an elliptical aperture with geometric parameters derived from the isophotal
analysis (see Section~\ref{fig:isophotal_analysis}). We used the {\sc pPXF} algorithm 
\citep{Cappellari2004,cappellari2017,Cappellari2023} to derive the line-of-sight (LOS) velocity distribution
(LOSVD) of both stars and gas in UDG\,6. The LOSVD is commonly parametrised via the LOS velocity 
(\Vlos), velocity dispersion (\Slos), and Gaussian-Hermite moments $h_3$ and $h_4$ 
\citep{Gerhard1993, vanderMarel1993}.  However, given the LSB nature
of UDGs and the limitations in the quality of the spectroscopic data (S/N $\sim$ 20), we can
constrain, at most, the first two moments of the LOSVD, focusing on the
optical spectral range ($4800$-$7000$\,\AA). In addition, since Balmer lines 
are clearly visible in the spectrum, we fitted the reddening \citep[see][for a description]{Cappellari2023}.

\begin{figure*}
    \centering
    \includegraphics[width=1.0\linewidth]{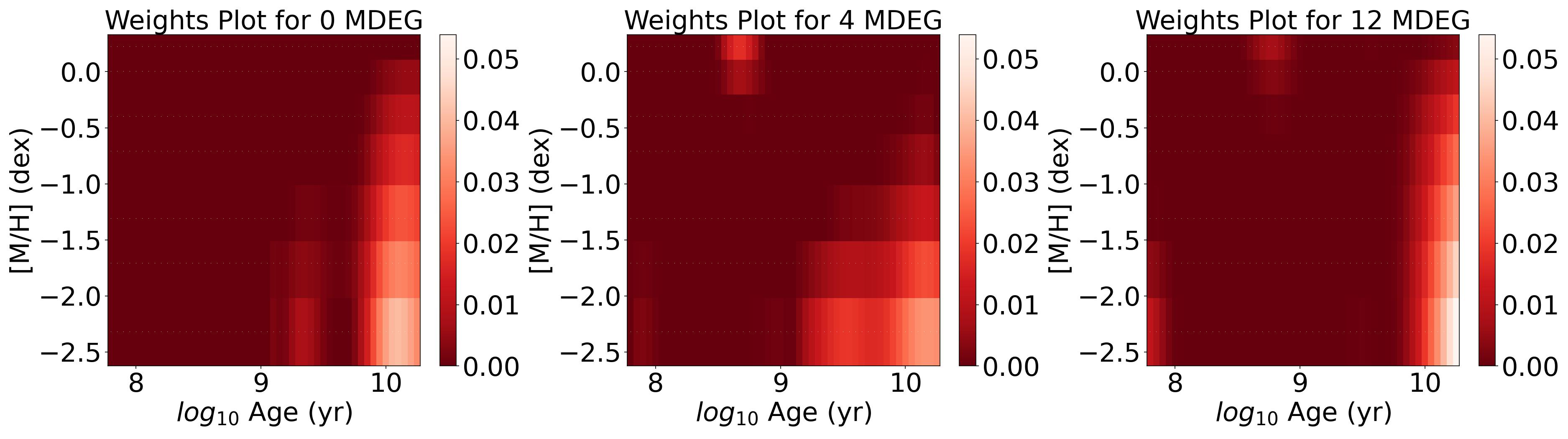}
    \caption{Stellar population analysis of 1D stacked spectrum of UDG\,6. Each panel shows the distributions of the stellar population light weights in the age-metallicity space for a spectral fit with multiplicative polynomials of degree equal to 0 (left), 4 (centre), and 12 (right).}
    \label{fig:Age_Z_panels}
\end{figure*}

We account for three components with a different LOSVD: one for the stars and 
two for the gas, the Balmer and the forbidden gas emission lines. For the stellar 
component, we adopted the E-MILES single stellar population (SSP) models 
\citep{Vazdekis2012, Vazdekis2016}, based on Padova isochrones \citep{Girardi2000} and 
a Kroupa \citep{Kroupa2001} initial-mass function. The E-MILES SSPs have a spectral 
resolution of FWHM$=2.51$\,\AA, and cover a wide spectral ($1680-50000$\,\AA),
metallicity ($-2.27\leq [M/{\rm H}]\leq+0.04$), and age 
($30\,{\rm Myr} \leq {\rm age} \leq 14\,{\rm Gyr}$) ranges. For the emission lines, 
we adopted four gas spectral templates, one for the Balmer lines (\ion{H}{$\alpha$} 
and \ion{H}{$\beta$}), fixing their ratio to the Balmer decrement, and the other 
three for the emission lines ([\ion{O}{III}], [\ion{N}{II}], and [\ion{S}{II}] doublets).  
The amplitude ratios of the emission lines are constrained by intrinsic theoretical
decrements, assuming no dust extinction (\ion{H}{$\alpha$}/\ion{\rm H}{$\beta$}$\,=2.86$, 
[\ion{S}{II}]$\lambda6716$/[\ion{S}{II}]$\lambda6731$\,$\in$\,[0.44, 1.43], 
[\ion{N}{II}]$\lambda6548$/[\ion{N}{II}]$\lambda6583$\,=\,0.33, [\ion{O}{III}]$\lambda4959$/[\ion{O}{III}]$\lambda5007$\,=\,0.33,
\citealt{Osterbrock2006}). As described in \PaperII, we used the MUSE 
line-spread function (LSF) of LEWIS data and adopted both multiplicative -- to correct the 
continuum shape for possible flux calibration imperfections in the spectrum
-- and additive -- to correct the mismatch between the stellar templates and the spectral 
continuum -- Legendre polynomials. We tested different values of the polynomial 
degrees in the range [0,12] with a step of 2, and evaluated the quality of the 
best-fit from the $\chi^2$ value. The best fit is obtained with both additive and
multiplicative polynomials of degree 4. 

The uncertainties on the fitted parameters have been derived using Monte 
Carlo simulations and following the same procedure as described in \PaperII. 
We generated 500 perturbations of the original spectrum by randomly adding 
Poissonian noise and removing/replacing portions of masked spectral regions,
without altering the final S/N of the perturbed spectra. The errors on \Vlos\, 
and \Slos\, are derived as the standard deviations of the distributions of 
the fitted kinematic parameters. 

In Fig.~\ref{fig:UDG6_bestfit}, we show the 1D stacked spectrum of 
UDG\,6 with its best-fit from {\sc pPXF}. From the best-fit of the gas component
(yellow lines), we obtained a value of \Vlos$_{\rm, gas} = 3559 \pm 1$\,km s$^{-1}$ for
all emission lines. We obtained values of \Slos\, consistent within 1$\sigma$ as well 
(\Slos$_{\rm, Balmer} = 18\pm1$\,km s$^{-1}$ and \Slos$_{\rm, forbidden} = 22\pm4$\,km s$^{-1}$). 
The spectral fit of the stellar component is instead poor. This is mainly due to
the absence of prominent absorption lines in the spectrum rather than the quality of
the spectrum (S/N $\sim$ 20). The spectral fit returned an exceptionally high value of
\Slos$_{, \ast}$, resulting in an artificial broadening of the best-fit profile (red line). 
Thus we can only rely on the fitted line-of-sight velocity, which is \Vlos$_{, \ast} = 3584\pm15$\,km s$^{-1}$. 
This value is consistent within 2$\sigma$ with the estimate obtained for the gas component.

\subsection{Age and metallicity of stars}
\label{sec:stellar_pop}

We used the spectral fitting algorithm {\sc pPXF} to derive the integrated stellar population 
properties of UDG\,6, following a similar approach as described in \cite{Doll2026} (hereafter \PaperV), fitting
the spectrum in the range $4800-5500$\,\AA, adopting E-MILES stellar templates 
and a regularisation factor of 50 (see \PaperV\,for a description). 
Since the value of velocity dispersion for the stellar component is unreliable, we 
extrapolate the $\sigma_{\rm eff}$ from the Faber-Jackson relation for dwarf galaxies
(\citealt{Kirby2013}, $\sigma_{\rm eff, FJ}=19$\,km s$^{-1}$), consistent 
with the \Slos$_{\rm, gas}$.
Given the extremely weak absorption features, the spectral fit 
is largely driven by the continuum shape, which can be significantly influenced 
by the adopted degree of the Legendre polynomials and by errors in the flux calibration. It is therefore essential 
to carefully assess the impact of the polynomial choice on the resulting fit.
To this aim, we tested the effect of different multiplicative degrees exploring 
values in the same range adopted for the kinematics. In Fig.~\ref{fig:Age_Z_panels}, 
we show the distributions of the stellar population light weights for three representative
cases: 0, which corresponds to the case of a constant multiplicative polynomial, 4, the same 
as the best-fit for kinematics, and 12, the largest value of the explored grid.

We noticed a significant fraction of weights located at age $\gtrsim3$\,Gyr with metallicity 
spanning in the range $[-2.5, -1$]\, dex, independently from the choice of the tested 
polynomial degree. However, the spectral fit does not allow to fully constrain the metallicity. 
A minor contribution of younger stellar populations
(< $10^8$ yr) is also present, but the weights that represent it are variable and negligible
compared to the older population both in terms of mass and of light contribution.
This younger population could be responsible for the presence of strong emission 
lines in the spectrum of UDG\,6. The best-fit with high polynomial degrees shows a 
consistent fraction of weights with nearly solar metallicity with an Age$~\sim3\cdot10^8-10^9$\,yr, which could be 
associated with the broad absorption line of \ion{H}{$\beta$}. To test our hypothesis, 
we masked this broad absorption feature and repeated the spectral fitting, adopting the 
same configuration of {\sc pPXF} and confirming that the solar-like metallicity
contribution disappeared. We additionally repeated the spectral fit by fixing 
the mean metallicity to different values in the range [$M$/H] $\in [0,-2]$\,dex
with a step of 0.5, covering the possible values of the MILES stellar templates
and investigating the quality of the resulting best-fit. However, all the fits 
share similar quality ($\Delta \chi^2\sim0.01$).
For the other multiplicative degree values 
explored in the range [2,12], the configurations of weights are similar to the 4 and 12 degree cases. 
We obtained similar results -- a persistent old-to-intermediate contribution and 
variable younger contributions for multiplicative degrees larger than 0 -- also for no 
regularization and for a regularization factor of 100.
We performed an additional test based on the predictions of \cite{Blakeslee2001b} and \cite{Vazdekis1996}, which relate stellar population age and metallicity to broadband colours. We constrained the colour within its uncertainty range ($g-r = 0.32 \pm 0.20$ mag) and required the stellar population age to be older than 3 Gyr. Nevertheless, the compatible models encompass a broad range of metallicities, preventing a robust constraint on this parameter.
We concluded that UDG\,6
contains a stellar component with an age of at least $\gtrsim3$\,Gyr, but its
metallicity cannot be constrained by spectral fitting.

\subsection{Star-forming regions}
\label{sec:star_forming_regions}

\begin{figure*}
    \centering
    \includegraphics[scale=0.62]{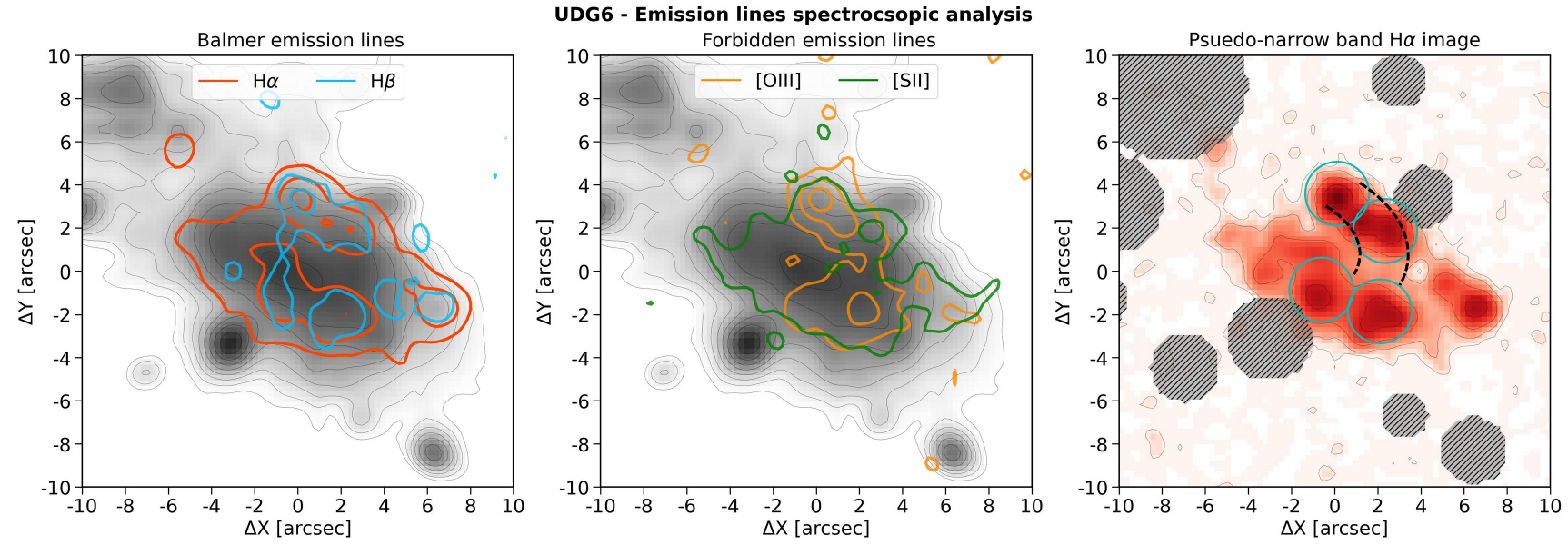}
    \caption{Emission lines analysis of UDG\,6. Left panel: the MUSE reconstructed image (grey) superimposed with contours of \ion{H}{$\alpha$} and \ion{H}{$\beta$} maps in red and blue, respectively. Central panel: the MUSE reconstructed image (grey) superimposed with contours of [\ion{O}{III}] and [\ion{S}{II}] maps in yellow and green, respectively. The contours of the MUSE image and the emission lines maps are convolved with a Gaussian kernel of size $1\times1$~pixel$^2$. Right panel: pseudo-narrow band of \ion{H}{$\alpha$} image. Grey hatched circles represent the masked regions, the cyan circles identify the apertures, and the black dashed line delimits the spiral-like feature.} 
    \label{fig:emission_line_maps}
\end{figure*}

We investigated the nature of the ionisation source of the emission lines in UDG\,6,
measuring specific emission lines ratios and building the \citet*{BPT1981} diagram 
(hereafter BPT). We identified the strongest lines in the galaxy spectrum
- \ion{H}{$\beta$}, the [\ion{O}{III}] doublet, \ion{H}{$\alpha$}, and the [\ion{S}{II}] 
doublet - and we constructed pseudo-narrow band images centred on these lines 
following the prescription presented in \cite{Hartke2025}. In the following, 
we summarise the procedure: we extracted $50$-pixel-wide band images centred 
on each emission line from the MUSE data-cube. The continuum was estimated from two
$20$-pixel-wide regions bracketing each line by computing the median flux, 
which was then subtracted to produce continuum-subtracted emission-line maps.
In Fig.~\ref{fig:emission_line_maps}, we show the maps of Balmer (left panel) 
and forbidden (right panel) emission lines. Contours of \ion{H}{$\alpha$} and \ion{H}{$\beta$}
are shown in red and blue, while those of [\ion{O}{III}] and [\ion{S}{II}] are
shown in yellow and green, respectively.

\begin{figure}
    \centering
    \includegraphics[scale=0.68, trim={1cm 0.1cm 1.5cm 1.5cm}, clip=True]{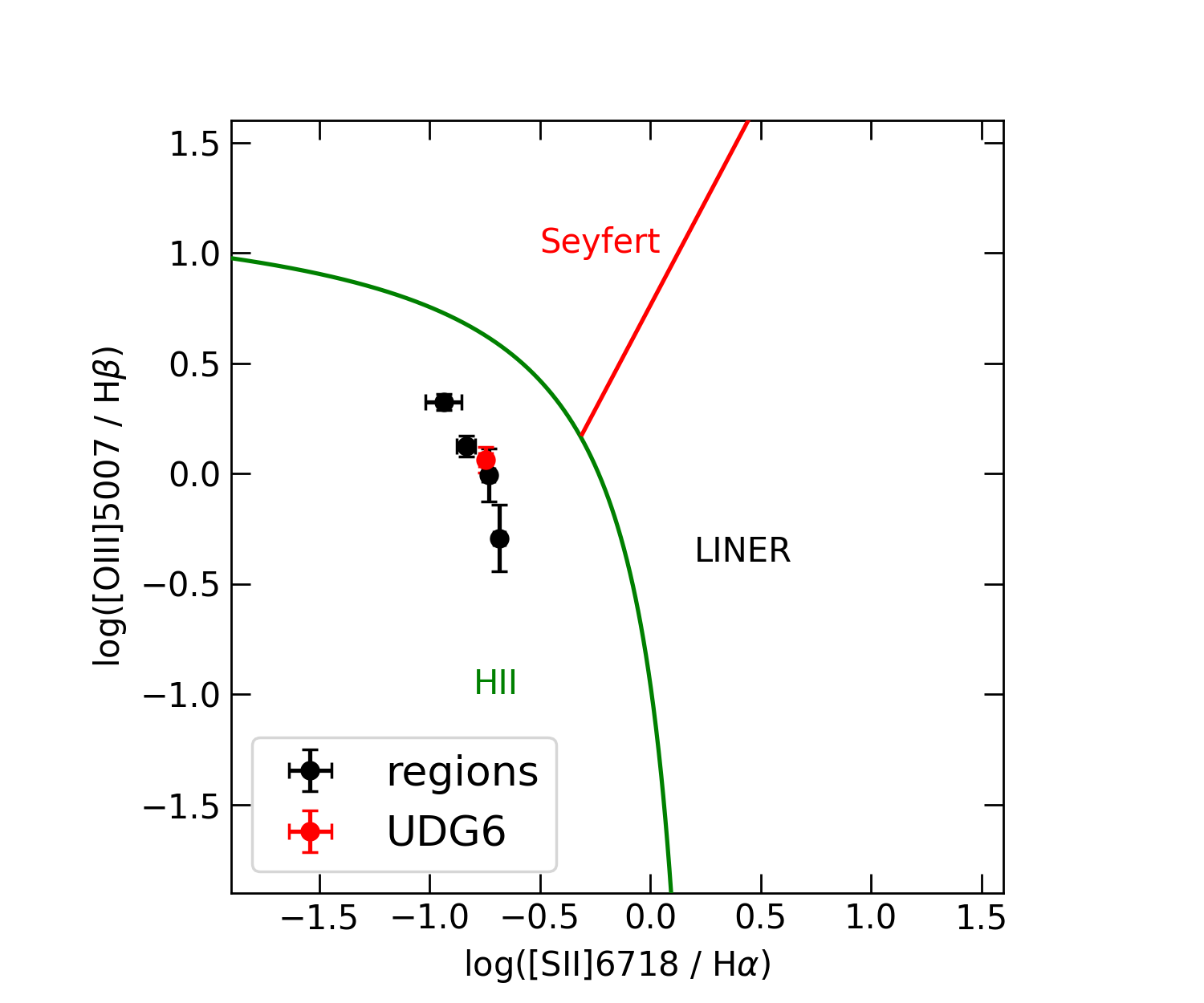}
    \caption{BPT diagram \citep{BPT1981}. The red circle is the value for UDG\,6 obtained from 1$R_{\rm eff}$ stacked spectrum, while the black circles mark the values obtained from the four apertures. The green and red lines separate the regions of the AGN galaxy, the LINER objects, and the ionised regions, respectively.} 
    \label{fig:BPT}
\end{figure}

The \ion{H}{$\alpha$} and [\ion{S}{II}] are widely distributed across the whole 
galaxy; in particular, the \ion{H}{$\alpha$} is the most abundant component. 
Its distribution presents multiple peaks and shows an extension in the north-east 
part of the galaxy. The distributions of [\ion{O}{III}] and \ion{H}{$\beta$} are 
instead more clumpy and patchy. We do not show the map for the emission line of 
the [\ion{N}{II}] since is too faint, though it was detected in the galaxy's stacked spectrum.

We corrected the flux of the emission lines for the Milky Way Galaxy 
extinction \citep{Fitz_extinction}, adopting as interstellar extinction and colour excess 
the values for the Milky Way ($A_V=0.217$ mag and $A_V/E(B-V)=3.1$) obtained 
from the NASA/IPAC Extragalactic Database\footnote{\href{https://ui.adsabs.harvard.edu/}
{https://ui.adsabs.harvard.edu/}} in the direction of Hydra I. 
We fitted the emission lines with a Gaussian function and derived the line 
strength by computing the area under the best-fitting profile. The uncertainties were estimated 
by propagating the standard deviations of the Gaussian fit parameters.
We extracted the emission line ratios [\ion{O}{III}]/\ion{H}{$\beta$} and 
[\ion{S}{II}]/\ion{H}{$\alpha$} from the 1D stacked spectrum and from 
four different regions across the galaxy extension, co-spatial with the major
emission peaks of \ion{H}{$\alpha$} and with radius equal to 1\,arcsec (Fig.~\ref{fig:emission_line_maps}).
In Fig.~\ref{fig:BPT} we show the BPT diagram for UDG\,6 $1\,R_{\rm eff}$-stacked spectrum
(in red) and for the other analysed regions (in black). The obtained values confirmed 
that UDG\,6 hosts several star-forming regions.

\subsection{Gas properties}
\label{sec:gas_pop}

The amplitude of gas emission lines can give us additional information on the nature
of ionised gas and on the star formation activity in UDG\,6. Before computing some useful emission 
line ratios, we estimated the galaxy internal extinction in $V$ band ($A_V$) by computing the 
Balmer decrement and adopting the \cite{Calzetti2000} attenuation law as implemented in \cite{Dominguez2013}:

\begin{figure*}
    \centering
    \includegraphics[scale=0.3]{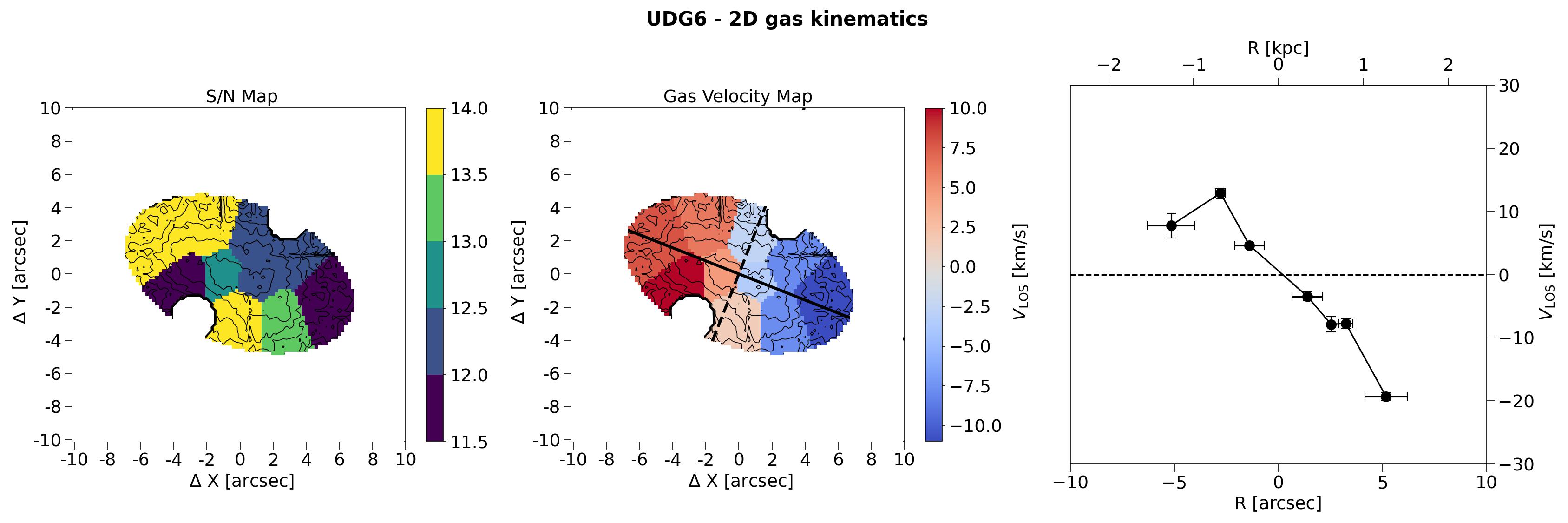}
    \caption{Spatially-resolved gas kinematics of UDG\,6. Left panel: Voronoi-binned map of the S/N. Central panel: ionised gas velocity map subtracted from the systemic velocity $V_{\rm syst}$. The black solid and dashed lines mark the photometric major and minor axes of UDG\,6, respectively. Right panel: velocity profile extracted along the galaxy's major axis. The radial distance $R$ is shown in projected angular (arcsec) and physical (kpc) distance.}
    \label{fig:gas_velocity_field}
\end{figure*}

\begin{equation}
\nonumber
A_V = (4.05 \pm 0.80) \times 1.97\log_{10}\left[\frac{[{\rm H\alpha/H\beta}]}{2.86}\right].
\end{equation}

\noindent We obtained a value of $A_V=1.2 \pm 0.5$ mag. The uncertainty on the reddening and all 
derived quantities are estimated via the error propagation standard formula.
We further computed the electron density parameter ($n_{\rm e}$), which gives us hints on the
compactness of star-forming regions. To this aim, we measured the flux ratio of [\ion{S}{II}] 
lines ($R$) and adopted the relation described in \cite{Proxauf2014}:

\begin{equation}
\begin{split}
        \log_{10}(n_{\rm e}) = 0.0543\,\tan(-3.0553R + 2.8506)\,+ \\
    6.98 - 10.6905R + 9.9186R^2 - 3.5442R^3,
\end{split}
\end{equation}

\noindent The relation provided in \cite{Proxauf2014} presents a convenient polynomial fit 
description to the classic statistical equilibrium equations and has been widely applied for \ion{H}{II}
diagnostics at low redshift. We thus obtained $\log_{10}(n_{\rm e})= 2.2 \pm 0.3\,{\rm cm}^{-3}$.

We estimated the gas metallicity following the calibration described in \cite{Dopita2016}, 
based on the N2S2H$\alpha$ line ratio. This calibration is valid in the electron density 
range $100 \leq n_{\rm e} \leq 400\,{\rm cm}^{-3}$ and our value falls within this range 
($n_{\rm e}\approx160\,{\rm cm}^{-3}$). The relation has also been successfully applied by \citealt{Lara-Lopez2022} on a sample of 
star-forming galaxies in the Fornax cluster. We obtained $12+\log_{10}({\rm O/H}) = 7.7 \pm 0.2$ dex.

Finally, we estimated the star formation rate (SFR) by computing: 
$ {\rm SFR}\, [{\rm M}_\odot/{\rm yr}]= 7.9 \times 10^{-42} L_{\alpha}$ \citep{Kennicutt1998}.
The parameter $L_{\alpha}$ corresponds to the intrinsic luminosity of 
\ion{H}{$\alpha$} corrected for dust extinction. We obtained a 
value for the SFR in UDG\,6 of $(1.9 \pm 0.1)\cdot10^{-3}\,{\rm M}_{\odot}$ 
yr$^{-1}$ and for the specific SFR of sSFR = SFR/$M_\ast = (7 \pm 3)\cdot 10^{-11}\,{\rm yr}^{-1}$.
In the estimation of the SFR, we did not explicitly account for a possible obscured star formation 
component, despite the high dust content of the galaxy. To investigate this, we computed the $V - R$ colour
map by extracting broadband images from the MUSE data-cube. If highly dust-obscured star-forming regions are
present, we expect to detect significantly redder colours and strong variations across the map. Instead, we 
find that the colours are of the order of 0.1 mag and show no significant spatial variation 
($\Delta (V - R) \sim 0.01$\,mag). Moreover, \ion{H}{$\alpha$} emission is broadly detected across the 
full extent of UDG\,6. The absence of spatially localised regions with anomalously red colours without 
\ion{H}{$\alpha$} detection argues against the presence of a significant amount of obscured star-forming regions.

We finally obtained an estimate of the mass of the ionised gas 
($M_{\rm gas}$ = $(9 \pm 2) \cdot 10^{6}$ M$_{\odot}$) following the prescription 
in \cite{gas_mass}. The final fraction of ionised gas mass over the total baryonic 
one ($M_{\rm b}=M_{\rm gas} +M_\ast$) is $f_{\rm gas}=M_{\rm gas}/M_{\rm b} = 0.24 \pm 0.08$. We did not consider any other
contribution of neutral hydrogen in the gas mass budget. Despite Hydra I having been covered
by the WALLABY@ASKAP survey \citep{Westmeier2022}, the HI mass sensitivity at 5$\sigma$ 
allows for the detection of $M_{\rm HI} \gtrsim 10^8 {\rm M}_\odot$. We therefore cannot exclude
the presence of HI mass lower than this limit in UDG\,6.
 All of the previously stated quantities are extracted from the
1\,$R_{\rm eff}$-stacked spectrum, representing thus a lower limit.

\subsection{Gas velocity field}
\label{sec:gas_vel_map}

We extracted spatially resolved kinematic information of the ionised gas in UDG\,6 following 
the same strategy described in \PaperII. We started masking all the stars, background, 
and foreground sources in the FOV of the UDG, and we applied an adaptive Voronoi 
tessellation algorithm \citep{Cappellari2003} to bin the data and obtain a 
specific S/N per bin. The estimator of the S/N adopted in \PaperII\, has been 
applied to derive the average S/N of spectra dominated by absorption lines 
and with a smooth stellar continuum. The UDG\,6's spectra contain instead strong
emission lines for which the previously adopted estimator would not provide an
accurate estimate of the S/N. For this reason, we additionally considered a 
dedicated line-based S/N estimator together with the S/N of the continuum. 

The S/N is defined as the minimum of the continuum S/N and the quadratic 
sum of the S/Ns of each detected emission line, to ensure a robust estimate 
of the S/N across various spectral configurations. 
The signal of the line is estimated by integrating the flux of the continuum-subtracted
emission line, while the noise is measured as the root mean square of the flux in two adjacent
regions bracketing the line emission. The S/N of the continuum is evaluated locally
in the same symmetrical regions around each line. Once the new S/N estimator is 
included in the algorithm, we applied the Voronoi binning on UDG\,6's data-cube with a threshold of S/N\,=\,10. 
This results in 10 independent bins with good-quality spectra (S/N$\,\sim12-14$.)

We fitted each binned spectrum with the {\sc pPXF} algorithm following a similar strategy 
for the 1D stacked spectrum as described in Section~\ref{sec:kin_1D}, with the only 
difference that instead of applying a specific spectral mask to each bin, we filtered 
the noisier regions by applying a 2 sigma-clipping. We extracted the gas kinematics by
combining Balmer and forbidden lines into a single component, since, when taken as a single one, 
the Balmer and forbidden lines do not show evidence of different kinematics. The uncertainties
were estimated by using the same Monte Carlo approach described in Section~\ref{sec:kin_1D}, 
by generating perturbations of the original spectrum and by repeating the spectral fitting.
These uncertainties reflect the precision of the spectral fitting algorithm in 
the recovery of the fitted parameters, given the data quality. These statistical 
uncertainties range between 0.5-2\,km s$^{-1}$.

In Fig.~\ref{fig:gas_velocity_field}, we show S/N map of the Voronoi bins (left panel), 
the velocity field of the gas (central panel) and the velocity profile extracted along the
galaxy's axes (right panel). The velocity profiles along the major axis shows a clear 
gradient in velocity. The amplitude of the rotation curve is
roughly \Vlos\,$\sim$ 14\,km s$^{-1}$. Assuming a galaxy inclination of
$i=\arccos(1-\epsilon)=66^\circ$, we obtained a deprojected maximum velocity of
$V_{\rm max, depr} = V_{\rm max}/\sin(i) = 22$\,km s$^{-1}$ along the galaxy's major axis.
The peak velocity of UDG\,6 occurs at $\sim1.3\,R_{\rm eff}$, corresponding to the radius where the maximum rotation velocity is expected for a pure exponential disk ($\sim2.2\,R_{\rm d}$ with $R_{\rm d}$ the disk scale length)
assuming $R_{\rm eff}\sim1.7\,R_{\rm d}$ \citep{Freeman1970}.

\subsection{ Analysis of GCs}
\label{sec:GCs}

The identification of GCs associated with UDG\,6 was carried out following the methodology 
introduced in \citet{Mirabile25}, which integrates MUSE spectroscopy with deep wide-field optical 
and near-infrared imaging from OmegaCAM@VST and VIRCAM@VISTA. From the analysis of the 
spectra of all compact sources in the field of UDG\,6, we did not find the presence of any bound GC or 
intra-cluster GC. To inspect the population of faint GCs, for which the spectral S/N is too low (S/N\,$\leq2.5\,\AA^{-1}$), 
we exploited the wide wavelength coverage of the MUSE cube to create three images approximately equivalent 
to the SDSS $g$, $r$, $i$ bands, which are then analysed together with the deep VIRCAM $H-$band observations of the cluster. 
We obtained photometric and morphometric information for all sources in the $1'\times1'$ field around UDG\,6 
and selected GC candidates based on their shape, colour–colour properties, and magnitudes ($m_H \geq 22\ {\rm mag}$). 
The resulting photometric catalogue is contaminated by Milky Way stars and faint background galaxies. We therefore
first removed all sources spectroscopically identified as stars, background galaxies, or emission-line objects. 
The residual contamination was then treated using a statistical background decontamination technique 
\citep{Dabrusco2016,Cantiello2020, Mirabile2024}. First, we characterised the contaminant population by 
analysing regions in an annulus between 15 and 30 arcsec from the galaxy centre, to determine the level of
local background contamination. On-galaxy GC candidates were instead selected within $1.5\,R_{\rm eff}$. 
The number of GCs was then estimated as the difference between the on-galaxy GC-candidate density and the
off-galaxy background density multiplied for the area within $1.5\,R_{\rm eff}$), corrected for the 
completeness of our data. The result of this analysis indicates that UDG\,6 hosts no significant GC overdensity; 
the value ($N_\mathrm{GC} = 0.2 \pm 5.4$) is consistent with zero within uncertainties (Mirabile et al. in prep.).

\section{Summary and discussion}
\label{sec:discussion}

In this Section, we briefly summarise the results obtained from the 
analysis carried out from the MUSE IF spectroscopy of UDG\,6 and their 
physical implications. Table~\ref{tab:properties_MUSE} reports the gas and
stellar properties of UDG\,6 derived in this work.

- Morphology: the isophotal analysis showed that UDG\,6 is characterised 
by an elongated structure with coherent orientation projected on the sky plane 
(Section~\ref{sec:isophotal_analysis}). The unsharp mask analysis revealed an 
arc-like structure on the west side of the main body of the galaxy (Section~\ref{sec:unsharp_mask}).

- Spectroscopic analysis: the UDG\,6's spectrum is characterised by the presence
of strong emission line of ionised gas, such as Balmer (\ion{H}{$\alpha$},
\ion{H}{$\beta$}) and forbidden ([\ion{O}{III}], [\ion{N}{II}], [\ion{S}{II}]) lines,
and extremely faint absorption features, such as \ion{H}{$\beta$}, MgI and 
iron lines. The spectral fitting analysis on the 1$R_{\rm eff}$-stacked spectrum 
revealed the co-spatial presence of a gas and stellar component, ensuring that the gas 
is not merely projected along the line of sight.
Due to the lack of strong absorption features, it was not possible to 
constrain the \Slos$_{, \ast}$ of the stellar component (Section~\ref{sec:kin_1D}). 

- Stellar populations: the lack of strong absorption lines prevented us 
from obtaining a tight constraint on the age and metallicity of the stellar component 
(Section~\ref{sec:stellar_pop}). The spectral fitting revealed a persistent old stellar component
($\gtrsim 3\,$Gyr). Unfortunately, we cannot constrain the stellar metallicity in UDG\,6 via spectral fitting.

- Emission line source: we recovered the nature of ionisation and spatial distribution
of the gas emission in UDG\,6 (Section~\ref{sec:star_forming_regions}). We found that \ion{H}{$\alpha$}
is the most abundant component and, together with [\ion{S}{II}], is widely distributed across the whole
galaxy's extension. The distributions of the \ion{H}{$\beta$} and [\ion{O}{III}] emissions 
appeared instead more clumpy and patchy. From the analysis of the BPT diagram, 
we confirmed that UDG\,6 is characterised by recent star-formation activity.

- Gas properties: UDG\,6 is characterised by a significant dust content ($A_V\sim1.2$\,mag),
diffuse star-forming regions ($log_{10}{(n_{\rm e})}\sim2$\,cm$^{-3}$) and a significant metal-poor
($12+\log_{10}{\rm (O/H)}=7.7$\,dex) ionised gas amount. The extremely low value for SFR
suggests that the star formation is inefficient or has been recently triggered 
(Section~\ref{sec:gas_pop}). The ionised gas fraction enclosed in $1\,R_{\rm eff}$ is $f_{\rm gas}=0.24\,\pm\,0.08$.

- Gas velocity field: we retrieved the gas velocity field of UDG\,6 and extracted the 
rotation curve along the major photometric axis of the galaxy. UDG\,6 shows clear hints of 
coherent rotation: Along the galaxy's major axis, the velocity curve reaches a maximum value 
of $V_{\rm max, depr}=22$\,km s$^{-1}$ on the deprojected galaxy plane (Section~\ref{sec:gas_vel_map}).

- GCs populations: the number of GCs in UDG\,6 is consistent with zero 
($N_\mathrm{GC} = 0.2 \pm 5.4$, Section~\ref{sec:GCs}).

Taking all these elements together, we can conclude that UDG\,6 shows evidence of an
arc-like underlying structure, possibly related to the ionised gas and due to an external perturbation.
UDG\,6 seems to have an old underlying stellar component, but its precise age and metallicity 
cannot be strongly constrained. UDG\,6 hosts a significant dust content and ionised gas amount with low
metallicity and a coherent velocity field that slowly rotates. The ongoing star-formation activity 
in the galaxy appears bursty and diffuse in its whole extension. In the following sections, 
we consider the global properties of UDG\,6 in relation to its location in the cluster and 
compare them with similar cases in the literature.

\begin{table}
\centering
\caption{Stellar and gas properties of UDG\,6.} 
\renewcommand{\tabcolsep}{0.25cm}
\renewcommand{\arraystretch}{1.}
\begin{tabular}{l l l  c }
\hline
\hline
     & Property                      &                     & UDG\,6               \\
\hline
(1)  & $M_{\ast, H}$                 & [$10^8 {\rm M}_\odot$]    &   0.18 $\pm$ 0.07    \\
(2)  & $V_{\rm LOS, gas}$            & [km s$^{-1}$]              &   3559 $\pm$ 1       \\ 
(3)  & $V_{\rm LOS, \ast}$           & [km s$^{-1}$]              &   3584 $\pm$ 15      \\
(4)  & $\sigma_{\rm LOS, Balmer}$    & [km s$^{-1}$]              &     18 $\pm$ 1       \\ 
(5)  & $\sigma_{\rm LOS, Forbidden}$ & [km s$^{-1}$]              &     22 $\pm$ 4       \\ 
(6)  & 12 + log$_{10}$(O/H)          & [dex]               &         7.7 $\pm$ 0.2         \\
(7)  & [$M$/H]                         & [dex]               &        $-1.4 \pm 0.2$         \\
(8)  & SFR                           & [10$^{-3}$ ${\rm M}_\odot\,{\rm yr}^{-1}$] &   $1.9 \pm 0.1$  \\
(9)  & sSFR           \rm                & [10$^{-11}$ ${\rm yr}^{-1}$]         &  $7 \pm 3$ \\
(10) & $M_{\rm gas}$                     & [10$^6$ ${\rm M}_\odot$]         &  $9 \pm 2$    \\          
(11) & $f_{\rm gas}$                     &                     & $0.24\pm 0.08$     \\
\hline 
\end{tabular}
\tablefoot{Properties of UDG\,6. (1): stellar mass derived from $H$-band image. (2-5): stellar and gas kinematics derived from $1\,R_{\rm eff}$-stacked spectrum. (6): gas metallicity. (7) stellar metallicity, derived by converting the gas metallicity with the prescription in \cite{Fraser-McKelvie2022}. (8-9): star-formation rate and specific star-formation rate (10-11) ionised gas mass and mass fraction enclosed in an aperture of $1\,R_{\rm eff}$.}
\label{tab:properties_MUSE}
\end{table}

\subsection{Is UDG\,6 a peculiar galaxy in Hydra I?}
\label{subsec:UDG6_environment}

\begin{figure}
    \centering
    \includegraphics[scale=0.42]{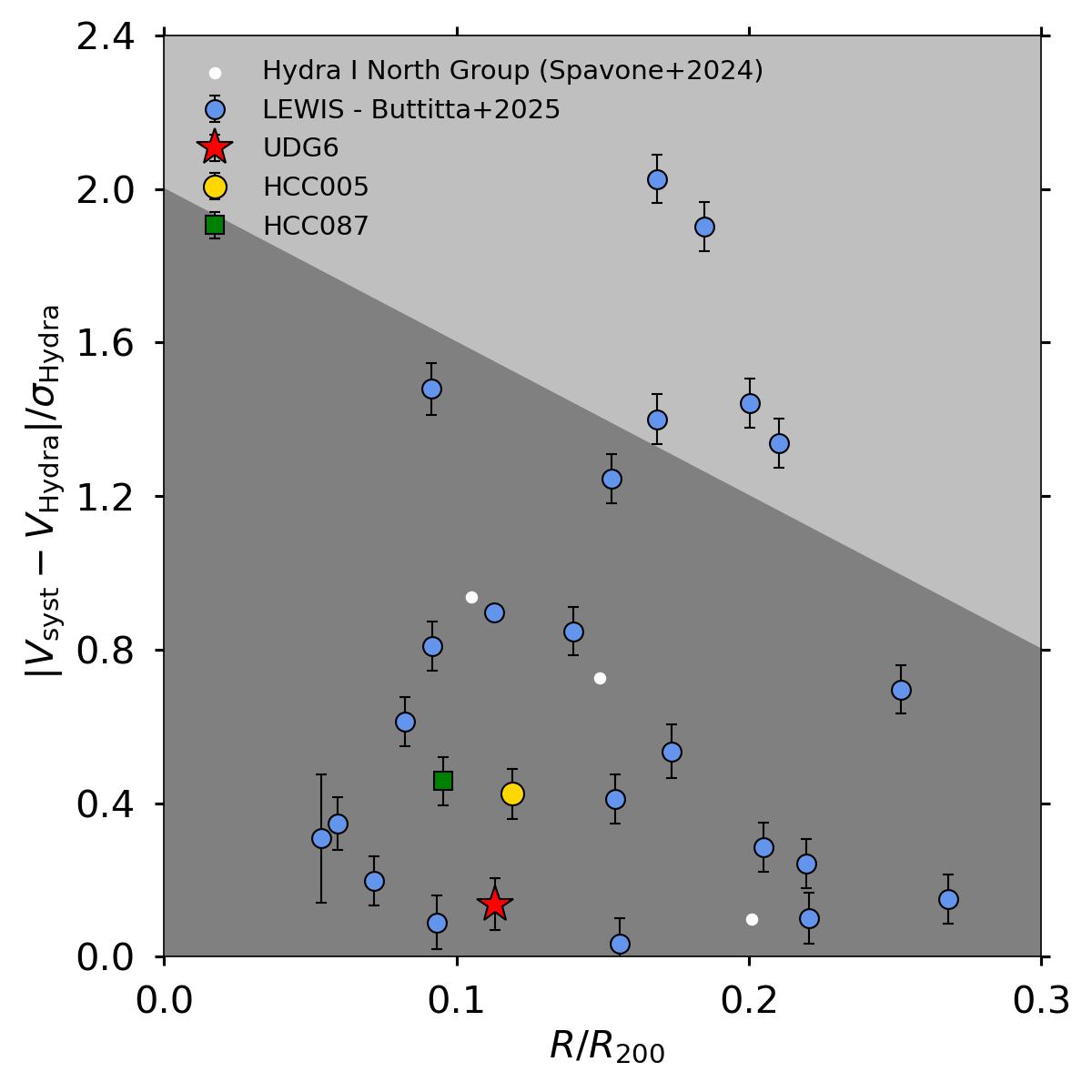}
    \caption{Projected phase-space diagram for galaxies in Hydra I. The galaxy's systemic velocity relative to the average velocity of the cluster, normalised by the cluster velocity dispersion ($|V_{\rm syst} - V_{\rm Hydra} |/\sigma_{\rm Hydra}$) is shown as a function of the projected clustercentric distance normalised by the cluster's virial radius ($R/R_{200}$). The dark and light grey regions represent the very early and the late infall regions, respectively. White and light blue dots represent the galaxies located in the North overdensity \citep{Spavone2024} and the LEWIS sample \citep{Buttitta2025}, respectively. The red star, yellow circle, and green square symbols show the position of UDG\,6, lenticular galaxy HCC\,005, and the disrupted dwarf galaxy HCC\,087, respectively.}
    \label{fig:phase-space}
\end{figure}

The projected systemic velocity of UDG\,6 (\Vlos$_{, \ast} = 3584 \pm 15$ km s$^{-1}$) is consistent with the $1\sigma_{\rm Hydra}$ velocity
distribution of the cluster ($V_{\rm Hydra} = 3683~\pm~46$\,~km s$^{-1}$, \citealt{Christlein2003}) 
where $\sigma_{\rm Hydra} = 724~\pm~31$\,~km s$^{-1}$, \citep{Lima-dias2021}. 
We can thus confirm that UDG\,6 is a cluster member of Hydra I, as the majority 
of the LEWIS sample (\PaperII). According to its relative position and velocity
in the projected phase-space diagram \citep{Forbes2023, Buttitta2025}, UDG\,6 
is located in the early infall region (Fig.~\ref{fig:phase-space}). 
This region contains the ancient infallers, i.e. galaxies that entered
the cluster at earlier epochs and are now virialised with the system.
There exists, however, the chance that galaxies classified as early infallers 
are instead interlopers ($\sim$10\%) or late infallers ($\sim$15\%) due to projection effects \citep{Rhee2017}.

In Hydra I, UDG\,6 is located in the Northern overdensity of the cluster \citep{LaMarca2022b}.
This region is characterised by strong evidence of galaxy interactions and by the
presence of intra-cluster light \citep{Spavone2024}. UDG\,6 lies at a projected distance
of 1.87\,arcmin from the lenticular bright galaxy HCC\,005, which corresponds to a projected 
separation of $\sim27\,$kpc (see Fig.~\ref{fig:UDG6_VST}). The relative velocity between
HCC\,005 (\Vsyst$\,=\, 3370\,\pm\,8$\,km s$^{-1}$) and UDG\,6 is $\Delta V\,\sim\,190$\,km s$^{-1}$. HCC\,005 shows 
two prominent tails, evidence of some environmental perturbation acting on the
galaxy. Nearby HCC\,005, in the south-eastern direction, is located a tidally disrupted
dwarf galaxy with a peculiar S-shape, HCC\,087 \citep{Misgeld2008, Koch2012}, a clear example
of on-going interaction. All the discussed cases suggest that the environment might
also have an impact on the structure of UDG\,6, inducing the formation of the arc-like
structure and triggered clumpy star-formation activity. 
A similar spiral-like structure has been reported in \cite{Mancera-Pina2024}, in a HI gas-rich UDG. However, there are important differences between them. UDG\,6 shows an asymmetric single-armed feature, whereas the galaxy presented by \cite{Mancera-Pina2024} appears symmetric and has a global disk-like spiral structure. Moreover, the two galaxies reside in different environments: UDG6 is located in the inner region of a cluster, whereas AGC\,114905 is found in an isolated/low-density environment. The asymmetric morphology of UDG6, combined with its location in the cluster and proximity to HCC\,005, supports an external origin, whereas the spiral structure in AGC\,114905 is consistent with internal disk processes in a gas-rich system.

The majority of the LEWIS galaxies ($\sim75\%$) are located in very early infall region
(\PaperII), which includes UDGs with a broad range of properties in terms of stellar
kinematics, metallicity, and star formation history (\PaperII\, and \PaperV). 
The UDGs of the LEWIS sample, similarly to typical UDGs in other clusters, are 
red, gas-poor, and quenched,  with dwarf-like or slightly higher metallicity. 
Most of these properties, constrained by the analysis of the stellar component, 
cannot be compared with UDG\,6's properties, because we do not have reliable results. 
In contrast, UDG\,6 seems to retain a significant amount of gas reservoir and presents 
hints of ongoing star formation activity. UDG\,6 may represent a rare example of a UDG
located in the virialised region of the cluster with evidence of ionised gas, or be an 
interloper which is starting to feel the cluster forces, thus leaving an unperturbed gas velocity field.

\begin{figure}
    \centering
    \includegraphics[scale=0.42]{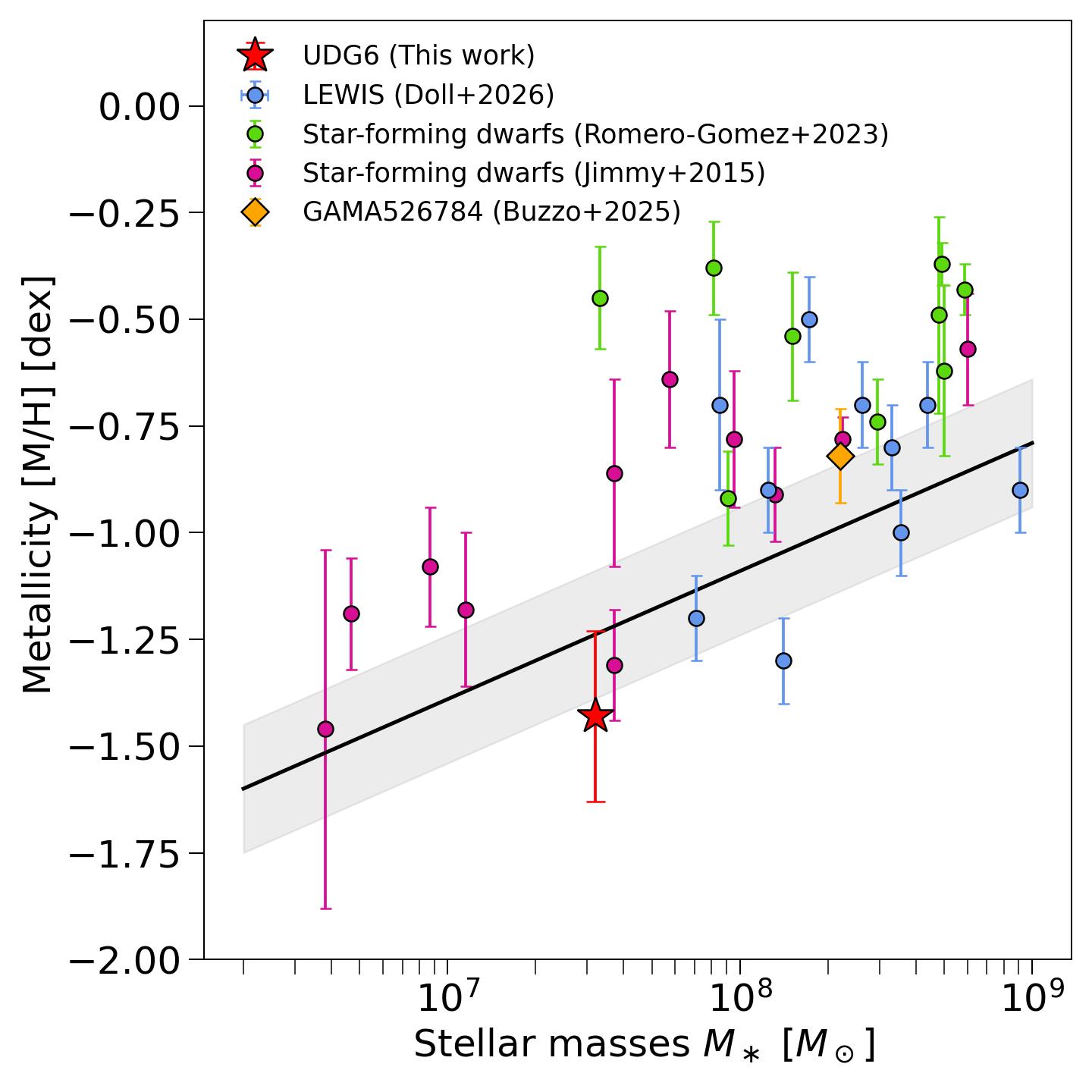}
    \caption{Mass-Metallicity relation. The black solid line and shaded region correspond to the mass-metallicity relation for dwarfs and $1\sigma$ uncertainties \citep{Kirby2013}. UDG\,6 is represented by the red star; its stellar metallicity was derived by converting the gas-phase measurement using the prescription in \cite{Fraser-McKelvie2022}. The LEWIS sample is marked with light blue dots \citep{Doll2026}. We additionally reported the value for GAMA\,526784 \citep{Buzzo2025} (orange diamond) and those for the samples of star-forming dwarf galaxies studied by \citealt{Jimmy2015} (magenta) and \citealt{Romero-Gomez2023} (green).}
    \label{fig:MZR}
\end{figure}

\subsection{A star-forming UDG in a cluster: a rare UDG or a transition phase?}  
\label{subsec:udgs_literature}

Despite a consistent number of gas-rich UDGs having been studied through HI data,
UDGs with strong signatures of ionised gas are rare. Recently, \cite{Khim2025} have discovered 
an unusual UDG with a coherent stellar velocity field, several GCs and star-forming knots characterised by 
clear \ion{H}{$\beta$}, [\ion{O}{II}] and [\ion{O}{III}] emission lines. Similarly, \cite{Buzzo2025}
combined HST and Subaru/HSC imaging data with IF spectroscopy from MUSE, to fully characterise a 
GC-rich star-forming UDG, where strong emission lines, such as Balmer lines 
(\ion{H}{$\alpha$} and \ion{H}{$\beta$}) and forbidden lines of [\ion{N}{II}], [\ion{S}{II}] and
[\ion{O}{III}], are detected, and a coherent stellar and ionised gas velocity field is derived. 
However, these UDGs are located in low-density environments. 

\cite{Kadowaki2017} reported the serendipitous discovery in the Coma Cluster of a UDG  
with evidence of ionised gas, characterised by [\ion{N}{II}], [\ion{O}{III}] 
and Balmer emission lines. \cite{Junais2021} presented a detailed multi-wavelength analysis of
a HI gas-rich UDG in the Virgo Cluster. Nearby this UDG, ionised-gas knots have been detected
in the same HI cloud associated with the UDG, suggesting that the latter has undergone some
ram-pressure stripping event, which removed HI gas clouds from it and subsequently triggered 
star-formation in clumpy knots. These UDGs, similarly to UDG\,6, are dominated by an 
old underlying stellar population. Their rarity suggests that the detection of ionised gas
in cluster UDGs could be a transient phenomenon, quickly quenched by the dense environment.
Recently, \cite{Taylor2026} reported the discovery of a gas-rich UDG in the Virgo cluster
whose \ion{H}{I} content has been displaced by ram-pressure stripping as the galaxy enters the 
cluster for the first time. Similar to UDG\,6, this system has suffered from environmental processes.
Environmental processes may also have affected the formation and evolution of the Hydra I UDG\,32,
which \cite{Hartke2025} propose to have formed from pre-processed material originating from the south-east group.

UDG\,6 hosts a significant dust content and it is characterised by an inefficient star-formation
activity with a SFR of $\sim2\cdot10^{-3}\,{\rm M}_\odot/{\rm yr}$. \cite{Kado-Fong2022} retrieved the SFRs of a 
sample of 22 nearby \ion{H}{I} rich UDGs located in low-density environments. Their estimates rely on an \cite{Kroupa2001} 
IMF, providing thus systematically lower values than those obtained with a \cite{Salpeter1955} IMF, and
were obtained from SED fitting including UV data, thus tracing different star formation timescales.
We convert the value of UDG\,6's SFR, to be comparable with the value obtained by \citealp{Kado-Fong2022} 
(${\rm SFR}_{\rm Kroupa} \approx {\rm SFR}_{\rm Salpeter}/ 1.5$), obtaining an even lower value 
(SFR$\,\sim1.3 \cdot 10^{-3}\,{\rm M}_\odot/{\rm yr}$). On average, the SFR values of the analysed sample are higher 
(SFR$\,\sim 10^{-2}-10^{-1}\,{\rm M}_\odot/{\rm yr}$) than the estimate retrieved for UDG\,6, with the only exception of the field UDG AGC\,219200, which shows a SFR comparable with UDG\,6's measure.
Slightly lower values have been reported in star-forming dwarfs \citep[$\sim 1.7\cdot10^{-4}\,{\rm M}_\odot/{\rm yr}$,][]{Lee2009}, 
whereas higher estimates have been found for field gas-rich LSB galaxies \citep[$2\cdot10^{-2}\,{\rm M}_\odot/{\rm yr}$,][]{Greco2018b}. 
Despite being somewhat consistent with literature values, the extremely low value of SFR can alternatively 
be explained by the stochasticity of the initial mass function \citep{IMF_stoc}, i.e. a discontinuous 
sampling of stellar mass distribution that can strongly bias the inferred SFR value. Finally, UDG\,6 hosts a 
discrete ionised gas content within its effective radius ($f_{\rm gas}=0.24\,\pm\,0.2$), characterised by a low gas metallicity 
($12+\log_{10}{\rm (O/H)}=7.7\,\pm\,0.2$\,dex). Since we were not able to constrain the metallicity of the stellar 
component from spectral fitting, we adopted the conversion proposed in \cite{Fraser-McKelvie2022} to obtain the stellar 
metallicity and compare it with similar studies. We therefore obtained [$M$/H]\,=\,$-1.43\,\pm\,0.2$\,dex.
Given its low stellar mass with respect to the other galaxies in LEWIS, UDG\,6 has the lowest 
metallicity of the sample, and it is consistent within $1\sigma$ with the \citealt{Kirby2013} 
relation (Fig.~\ref{fig:MZR}). Its metallicity is also consistent with 
the value of star-forming dwarf galaxies even down to stellar masses of
$M_\ast\sim10^{6.6}\,{\rm M}_\odot$ \citep{Jimmy2015}.

\subsection{On the formation channel of UDG\,6}
\label{subsec:formation_channels}

Considering all the quantities obtained from spectroscopic analysis, we review the
various formation scenarios proposed for UDGs with a discrete gas amount by attempting to identify
the most suitable for UDG\,6. Gas-rich UDGs might originate from `puffed-up dwarfs', whose stellar
distribution is stretched to larger radii, while retaining significant amounts of gas. The involved
mechanisms could be either internal or external. High-spin DM halos or frequent episodes of star-formation
feedback push the gas from the galaxy's centre to the outskirts, thus transforming a dwarf galaxy 
into a diffuse system and suppressing the star-formation activity \citep{Amorisco2016, Rong2017, DiCintio2017}. 
Alternatively, UDGs with a moderate-to-low amount of gas can originate from weak gravitational interaction 
and galaxy mergers, which can inflate dwarf galaxies into UDG-like systems \citep{Bennet2018, Muller2019}. 
Gas-rich UDGs might form from collisional debris from galaxy mergers or close encounters 
\citep{Lelli2015, Duc2014, Ploeckinger2018, Ivleva2024} or from gas clumps from ram-pressure 
stripped galaxies infalling into the clusters \citep{Poggianti2017, Grishin2021}.

Despite not being precisely constrained in age and metallicity, UDG\,6 shows evidence of an old 
underlying stellar population. The distribution of stars and ionised gas is nearly co-spatial and almost 
uniformly distributed in the galaxy's extension. In addition, the low-metallicity gas
coherently rotates within the galaxy. These features indicate that the star formation activity 
has been suppressed, preventing the enrichment of the gas and supporting the idea that UDG\,6 
has evolved as a normal dwarf galaxy with a significant amount of gas before becoming diffuse.
Together with the morphological structure and coherent gas rotation, UDG\,6's properties disfavour 
the scenario in which UDGs form via internal mechanisms, which tends to remove all the gas from the
innermost region of the galaxy. We can further exclude the ram-pressure stripping scenario, 
since the UDGs originated with this mechanism would have an irregular morphology and a dominant young
stellar component.

A reasonable evolutionary path for UDG\,6 seems to be the `puffed-up scenario', similarly to the
other UDGs in the LEWIS sample (\PaperV), but due to external processes. A strong interaction with the
hot intra-cluster medium would have stripped the gas from UDG\,6, strongly altering its coherent rotation. 
Most probably a weak tidal interaction with nearby massive galaxies as proposed by \cite{Bennet2018} 
can originate UDG-like system with properties similar to UDG\,6. A mild environmental perturbation from 
the cluster, possibly due to the dynamically active region where UDG\,6 resides, may have contributed to 
triggering only recently a clumpy star-formation activity that we observed today. This would justify 
the extremely low value for SFR and the unperturbed gas with a coherent rotation.

\section{Conclusions}
\label{sec:conclusions}
We presented a detailed spectroscopic analysis of a gas-rich UDG in the Hydra I cluster,
identified within the LEWIS project, showing clear evidence of diffuse star-forming regions.
This galaxy is the only object in our sample exhibiting strong emission lines in the MUSE 
data cube. For UDG\,6, we constrained the properties of the ionised gas, including its morphology, 
kinematics, and metallicity. However, the age and metallicity of the stellar component
remain poorly constrained; we can only infer the presence of a metal poor old-to-intermediate-age 
stellar population. We argued that UDG\,6 could be in a transient evolutionary phase, in 
which a quenched UDG, originating from a puffed-up dwarf via external mechanisms, can 
re-ignite localised star formation under environmental effects.

Further insight into the nature and evolutionary pathway of UDG\,6 will require a multi-wavelength 
approach. Ultraviolet observations will help constrain the recent star formation rate, 
while deep HI data will probe the presence of neutral gas reservoirs, providing a critical test
of the proposed scenario. In addition, combining multi-band photometry with spectral energy 
distribution (SED) fitting will enable more robust constraints on the underlying stellar 
populations. Retrieving an accurate stellar velocity dispersion will allow us to infer 
the DM content, a fundamental key parameter to disentangle the different proposed formation
 pathways for UDG\,6. All these quantities will allow a comprehensive characterisation of this 
peculiar UDG and clarify the role of the environment in shaping its properties.

These efforts will also be essential for mapping the UDG population at larger 
cluster-centric distances. Using new deep imaging covering the Hydra I cluster out 
to the virial radius, we have already identified 24 new UDG candidates (Borsato et al., in prep.).
Follow-up spectroscopic observations will improve the statistical characterisation of 
star-forming UDGs and their connection to the cluster environment.

\begin{acknowledgements}
We wish to thank the anonymous Referee whose comments helped us to improve the clarity of the manuscript. Based on observations collected at the European Southern Observatory under ESO programmes 108.222P.001, 108.222P.002, 108.222P.003. 
This work is based on the funding from the INAF through the GO large grant in 2022, to support the LEWIS data reduction and analysis (PI E. Iodice). The authors wish to thank P. Mancera Piña, O. M\"uller, E. Sola, and L. Coccato for the useful comments and discussions on the work presented in this paper. 
J.H acknowledges support from TCSMT through a starting grant. 
JF-B acknowledges support from the PID2022-140869NB-I00 grant from the Spanish Ministry of Science and Innovation. 
DF thanks the ARC for financial support via DP250101673. 
EMC recognises the support of the Italian Ministry of University and Research (MUR) grant PRIN 2022 2022383WFT ``SUNRISE'' (CUP C53D23000850006) and Padua University grants DOR 2023-2025. 

\end{acknowledgements}

\bibliographystyle{aa.bst}
\bibliography{LEWIS_paperVI.bib}

\end{document}